\documentclass[aps,prb,showpacs,reprint,usenames,dvipsnames]{revtex4-1}
\usepackage{graphicx,color,hyperref}
\usepackage[utf8]{inputenc}
\usepackage{etoolbox}
\usepackage{dcolumn}
\usepackage{bm}
\usepackage{amsmath}

\newcommand{\MnI}{Mn$_{i}$}
\newcommand{\MnBi}{Mn$_{\rm Bi}$}
\newcommand{\TeBi}{Te$_{\rm Bi}$}
\newcommand{\BiTe}{Bi$_{\rm Te}$}
\newcommand{\tcr}{}
\newcommand{\tcb}{}
\newcommand{\tcg}{}

\begin{document}

\title{Electronic and transport properties of the Mn-doped 
\tcg{topological insulator} Bi$_{2}$Te$_{3}$: A first-principles study}

\author{K. Carva$^{1}$, J. Kudrnovsk\'y$^{2}$, F. M\'aca$^{2}$, 
V. Drchal$^{2}$, I. Turek$^{1}$, P. Bal\'a\v{z}$^{1}$, 
V. Tk\'a\v{c}$^{1}$, V. Hol\'y$^{1}$, 
V. Sechovsk\'y$^{1}$ and J. Honolka$^{2}$}

\affiliation{$^{1}$Charles University in Prague, Faculty of Mathematics 
and Physics, Dept. Condensed Matter Physics, Ke Karlovu 5, 
CZ-12116 Prague 2, Czech Republic}

\affiliation{$^{2}$Institute of Physics, Academy of Sciences of the 
Czech Republic, Na Slovance 2, CZ-18221 Prague 8, Czech Republic}

\date{\today}

\begin{abstract}
We present a first-principles study of the electronic, magnetic,
and transport properties of the topological
insulator Bi$_{2}$Te$_{3}$ doped with Mn atoms in substitutional (\MnBi{}) and
interstitial van der Waals gap positions (\MnI{}), which act as acceptors 
and donors, respectively.
The effect of native \BiTe{}- and \TeBi{}-antisite defects and
their influence on calculated electronic transport properties is 
also investigated.
We have studied four models representing typical cases, namely,
(i) Bi$_{2}$Te$_{3}$ with and without native defects, (ii) \MnBi{} 
defects with and without native defects, (iii) the same, but for 
\MnI{} defects, and (iv) the combined presence of \MnBi{} and \MnI{}.
\tcb{It has been found that lattice relaxations around \MnBi{} defects play
an important role for both magnetic and transport properties}.

The resistivity is strongly influenced by the amount of carriers,
their type, and by the relative positions of the Mn-impurity energy levels
and the Fermi energy. \tcg{Our results suggest strategies to tune
bulk resistivities, and also clarify the location of Mn
atoms in samples. Calculations
indicate that at least two of the considered defects have to be present
simultaneously in order to explain the experimental observations,  and the role
of interstitials may be more important than expected.}

\end{abstract}


\maketitle

\section{Introduction}
\label{intro}

The field of spintronics is connected with a continuous search for new magnetic
semiconductors.
The most popular in the past were materials based of the group
III-V or II-VI semiconductors doped with magnetic elements, mostly
with manganese, (Ga,Mn)As being the prototypical case. \cite{r_14_Dietl_RMP_DilFMSC,r_10_Sato_Kud_Turek_FP_DMS}
Here, the main idea is to prepare \tcg{functional compounds where carrier concentrations and magnetic
degrees of freedom are entangled.} 

\begin{figure}
\includegraphics[width=\columnwidth]{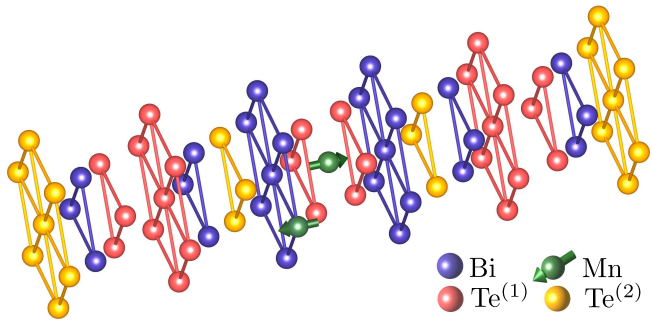}

\caption{\label{fig:struct} Crystal structure of the tetradymite Bi$_{2}$Te$_{3}$. A hexagonal supercell is shown. Possible Mn dopant positions, namely the substitutional one (\MnBi{}) and the interstitial one in the van der Waals gap (\MnI{}) are indicated. Note the two non-equivalent Te positions.}

\end{figure}

Another class of new materials has tetradymite crystal
structure like, e.g., Bi$_{2}$Te$_{3}$ or Bi$_{2}$Se$_{3}$, known
also for their thermoelectric applications. \cite{r_08_Noh_SOeff_Bi2Te3_ARPES}
It should be noted that Bi$_{2}$Te$_{3}$ or Bi$_{2}$Se$_{3}$ recently
attracted a great interest also due to the topological character of their electronic structure
which manifests itself in the existence of a Dirac surface state \tcg{with helical spin texture}. \cite{r_07_Fu_Kane_3DTI,r_09_Zhang_TI_BiTe_BiSe_SbTe_NPhys,r_10_Hasan_TI_RevMod,r_10_Zhang_FPstudy_BiTe_BiSe_SbTe_NJOP,r_09_Hsieh_ObservDiracCone_BiTe_SbTe}
Their doping by low concentrations of transition metal elements, e.g.,
by Ti, V, Cr, Mn, or Fe, \tcg{is a way to add novel
functionality to topological insulators.\cite{r_14_Dietl_RMP_DilFMSC}} It has led to ferromagnets with low Curie
temperatures \tcg{below} 20~K.\cite{r_10_Hor_Hasan_FM_TI_Mn-BiTe,r_05_Choi_Mn-doped_V_VI_SpGlass,r_14_Lee_Magnetism_transp_n-type_Mn-BiTe,r_16_Taras_cb_MagStruct_Mn_BiSe}
Theoretical calculations,\cite{r_14_Vergn_Mertig_ExchInt_BiSe_BiTe_SbTe} however, \tcg{obtain} a monotonic increase 
of the Curie temperature with increasing doping.
Doping by transition metal impurities leads to formation of carriers,
e.g., Mn-impurities in Bi$_{2}$Te$_{3}$ which substitute Bi-atoms
will act as acceptors. \tcg{Substitutional Mn positions (\MnBi{}) are assumed in most of the
works.\cite{r_10_Hor_Hasan_FM_TI_Mn-BiTe,r_12_Henk_TopChar_Dirac_Mn-BiTe,r_05_Choi_Mn-doped_V_VI_SpGlass,r_14_Lee_Magnetism_transp_n-type_Mn-BiTe} Contrary to this, in experiments many samples have exhibited electron-like conductivity,\cite{r_14_Lee_Magnetism_transp_n-type_Mn-BiTe,r_13_Watso_BiTeMn-Prop-Istit,r_15_Ruzicka_Holy_Mn-BiTe_properties} which indicates that \MnBi{} cannot be the only defect present.}
Similarly to (Ga,Mn)As, Mn atoms in Bi$_{2}$Te$_{3}$ can also
occupy interstitial positions, presumably an octahedral position with 6 nearest Te neighbours in the van der Waals (vdW) gap \cite{r_13_Watso_BiTeMn-Prop-Istit,r_15_Ruzicka_Holy_Mn-BiTe_properties} as shown in Fig. \ref{fig:struct}.
According to first principles calculations \cite{r_13_Zhang_MagDop_BiSe_BiTe_SbTe_FormEn} this position has the lowest formation energy of the three possible interstitial ones and its preference has also been found experimentally.\cite{r_15_Ruzicka_Holy_Mn-BiTe_properties} In such a case, Mn defects act as donors.
Related physical properties like the residual resistivity, 
magnetoresistance, or anomalous Hall effect, are of great importance
for the physical characterization of the doped Bi$_{2}$Te$_{3}$ or
Bi$_{2}$Se$_{3}$.

In addition to the above mentioned intentional doping by transition
metal defects there exist also native defects depending on growth
conditions, e.g., the Bi-rich or Te-poor samples in Bi$_{2}$Te$_{3}$.  
Such native antisite defects act as acceptors (in Bi-rich conditions) or
donors (in Bi-poor conditions), and thus play a key role in controlling bulk transport properties. \tcg{Both antisite defects are energetically more favourable than possible Te or Bi vacancies\cite{r_12_Scanlon_TIBulkCon_Antisite} in this compound. } 

\tcg{ Understanding bulk transport is of particular interest regarding the 2D topological surface state contributions to transport measurements. \cite{r_11_Hor_SC_SuppressBulkCond_Bi2X3,r_14_Hoefer_Cond_Via_SurfState_Bi2Te3,r_15_Brahl_TransportTI_Met2Ins}} The aim of the present work is thus a systematic first-principles
study of bulk transport properties of Mn-doped Bi$_{2}$Te$_{3}$
including the presence of native defects.
A prerequisite for such a study is a detailed understanding of
corresponding electronic structure properties as manifested,
e.g., in modifications of their densities of states (DOS), the
type of carriers and  the position of impurity levels with respect 
to the Fermi energy ($E_{\rm F}$), which is related to the strength 
of the impurity scattering.
So far, there is no related first-principles study for transport properties
of doped topological insulators in the literature, and also the study 
of alloy electronic structures are very rare.
Authors of existing studies limited their attention to the estimate
of formation energies of native \cite{r_12_Scanlon_TIBulkCon_Antisite,r_14_Oh_Lee_Antisite_BiTe_BiSe} and
transition metal \cite{r_13_Zhang_MagDop_BiSe_BiTe_SbTe_FormEn} defects using the supercell approach.
A combined study of both native and transition metal defects, in
particular for low doping concentrations, is technically too difficult
in the framework of the supercell approach.
A more suitable tool in such cases is the treatment of disorder in the framework of the coherent potential approximation \cite{r_67_ps}
(CPA), which allows to treat low concentrations of few defect types.
The CPA gives reliable concentration trends, but, e.g., the local
environment effects are beyond the scope of this method.
In this sense, the supercell and CPA methods are complementary. \tcg{Notably, experiments indicate no clustering of Mn atoms when samples are carefully grown. \cite{r_16_Taras_cb_MagStruct_Mn_BiSe} Such homogeneous distribution of dopants is an important condition for the CPA applicability.}
So far there exist two recent theoretical CPA studies \cite{r_12_Henk_TopChar_Dirac_Mn-BiTe,r_14_Vergn_Mertig_ExchInt_BiSe_BiTe_SbTe} of doped topological 
insulators including \MnBi{}-doped Bi$_{2}$Te$_{3}$, but \tcb{neglecting 
possible lattice relaxations}. \tcg{Most importantly, our CPA approach allows
a consistent treatment of transport properties using 
the linear response theory (the Kubo-Greenwood approach). For the first time we provide 
bulk residual resistivities of Mn-doped Bi$_{2}$Te$_{3}$, also with various native defects and 
both in the basal plane and along the $c$-axis. }

\section{Formalism}
\label{form}

The ideal Bi$_{2}$Te$_{3}$ consists \tcg{ of quintuple layers, which are separated by vdW gaps. Quintuple layers are composed from hexagonal layers following a Te(I)-Bi-Te(II)-Bi-Te(I) sequence, containing equivalent Bi layers and inequivalent center (Te(II)) and terminating Te(I) layers, see Fig. \ref{fig:struct}. }

The experimental lattice parameters \cite{r_10_Zhang_FPstudy_BiTe_BiSe_SbTe_NJOP}
 are used and we neglect
small volume changes due to Mn and antisite dopings.
\tcb{On the other hand, we include possible lattice relaxations 
around \MnBi{}, but neglect relaxations corresponding to native 
antisites.}

The electronic structure is studied using the Green function
version of the tight-binding linear muffin-tin orbital method\cite{r_84_Skriver_LMTO_book} with atomic-sphere approximation (TB-LMTO-ASA)
method, the effect of disorder is included in the CPA.
\cite{r_97_tdk}
The local spin-density approximation, the Vosko-Wilk-Nusair
exchange-correlation potential \cite{r_80_vwn} and the $s,p,d$-basis
are used.
Relativistic corrections were included approximately by adding
the on-site spin-orbit coupling term to the scalar-relativistic
TB-LMTO Hamiltonian and the value of the spin-orbit coupling
was determined self-consistently during calculations. \cite{r_08_tdk_LMTO_SO}
There are two reasons for the inclusion of the spin-orbit coupling:
(i) Te and in particular Bi atoms are heavy elements, and
(ii) the bandgap and its value is reproduced reasonably well as
compared to the scalar-relativistic case.
In the disordered case we have also used a simple screened impurity
model \cite{r_95_Korzh_Ruban_ScrImp_Madelung_CPA} to improve the treatment of the alloy electrostatics.
The empty sphere is used to simulate vdW 
positions in an ideal Bi$_{2}$Te$_{3}$ crystal\tcg{, which also improves the space filling}. 
The formula unit thus consists of six sublattices.
\MnBi{} atoms randomly occupy two equivalent Bi-sublattices in equal
amounts, while \MnI{} atoms can occupy randomly part of interstitial
positions.
Native defects are treated as \BiTe{} and \TeBi{} substitutionals
assuming the same concentrations on each of the corresponding sublattices.

The supercell VASP calculations were performed using the projector
augmented wave scheme\cite{r_99_Kresse_PAW} and both the Perdew-Zunger-Ceperly-Alder exchange
correlation potential,\cite{r_81_Perdew_Zunger_SIC_DFT,r_80_Ceperley_Alder_LDA_GS} and \tcg{the generalized gradient approximation \cite{r_96_Perdew_GGAsimple}} to simulate Mn defects of both
types and to investigate the role of lattice relaxations
around Mn atoms.
Supercells contained 60 atoms. \tcg{Mn impurities either substituted one Bi atom, or were placed in the vdW
gap position.} \tcg{During finalization of the paper, we became aware of a recent 
work\cite{r_14_Li_FM_SurfS_Mn-BiTe_DFT} based on VASP calculations, in which similar electronic structure
results for \MnBi{} and \MnI{} were obtained. }

\tcr{Calculations based on the supercell approach indicate that
a proper treatment of the size mismatch among Bi- and Mn-atoms
for \MnBi{} defects should be considered\tcg{, it leads to a lattice relaxation around the Mn atom}.}
In the TB-LMTO-CPA calculations we \tcg{ have used an already developed method 
\cite{r_90_kd_LMTO_CPA_diffWS,r_09_kmt_Redin_AFM_Fe_Ir001}
to include the size mismatch between Bi and Mn atoms 
treated naturally in the full potential approach \cite{r_03_Perss_Zunger_N_states_GaAsN}. In this way }the total magnetization and 
the important features of the DOS, in particular the relative positions of Mn levels with 
respect to $E_{\rm F}$, were reproduced to a good accuracy.
We refer the reader to \tcg{ the Supplemental Material\cite{r_Suppl-MnBiTe}  for details.} On the other hand, small \MnI{} atoms \tcg{can be modelled}
without modifications in the conventional TB-LMTO approach.
The supercell calculations obtain negligible lattice
relaxations in this case.

% \footnote{See Supplemental Material at [URL will be inserted by publisher] for details of the treatment of alloy disorder with unequal constituent atom radii}

As concerns transport properties, we limit ourselves to the
diagonal elements of the conductivity tensor to estimate
residual resistivities.
The formalism for diagonal elements of the conductivity tensor
including spin-orbit coupling \cite{r_10_t_Zalezak_ReRe_CoNi_CuNi} has the form identical
to that obtained in the scalar-relativistic case. \cite{r_02_tkd}
The present implementation employs a non-random velocity operator
formulation \cite{r_02_tkd}, which simplifies the evaluation of
disorder-induced vertex corrections. \cite{r_06_ctkb_vertex}
The estimate of the off-diagonal elements of the conductivity
tensor related, e.g., to the anomalous Hall effect, although possible
in the present formalism, is beyond the scope of the present study.

We have considered the following models for both the electronic structure 
and residual resistivities:
(i) Model A: Bi$_{2}$Te$_{3}$ without and with native acceptors and donors;
(ii) Model B: Bi$_{2}$Te$_{3}$ with \MnBi{} dopants as well as with native
acceptors and donors;
(iii) Model C: Bi$_{2}$Te$_{3}$ with Mn interstitials (\MnI{}) in the vdW
gap position as well as with native acceptors and donors; and
(iv) Model D: Combined effect of \MnBi{} and \MnI{} dopants.

\section{Results and discussion}
\label{res}

\subsection{Electronic structure}
\label{elstr}

Below we present the DOS's for studied Models A, B, C, D calculated by the TB-LMTO method.
We assume the same concentration $x$ of Mn dopants per formula 
unit in Models B, C, D, i.e., $x$=0.05.
Discussion of the DOS's, namely the amount and the type of carriers and 
the relative position of Mn-impurity peaks with respect to $E_{\rm F}$ 
allows to draw qualitative conclusions about corresponding 
resistivities shown below in Table~\ref{tab_rho}.

\subsubsection{Model A: Bi$_{2}$Te$_{3}$ without and with antisite 
defects}
\label{modA}

\begin{figure}
\includegraphics[width=9cm] {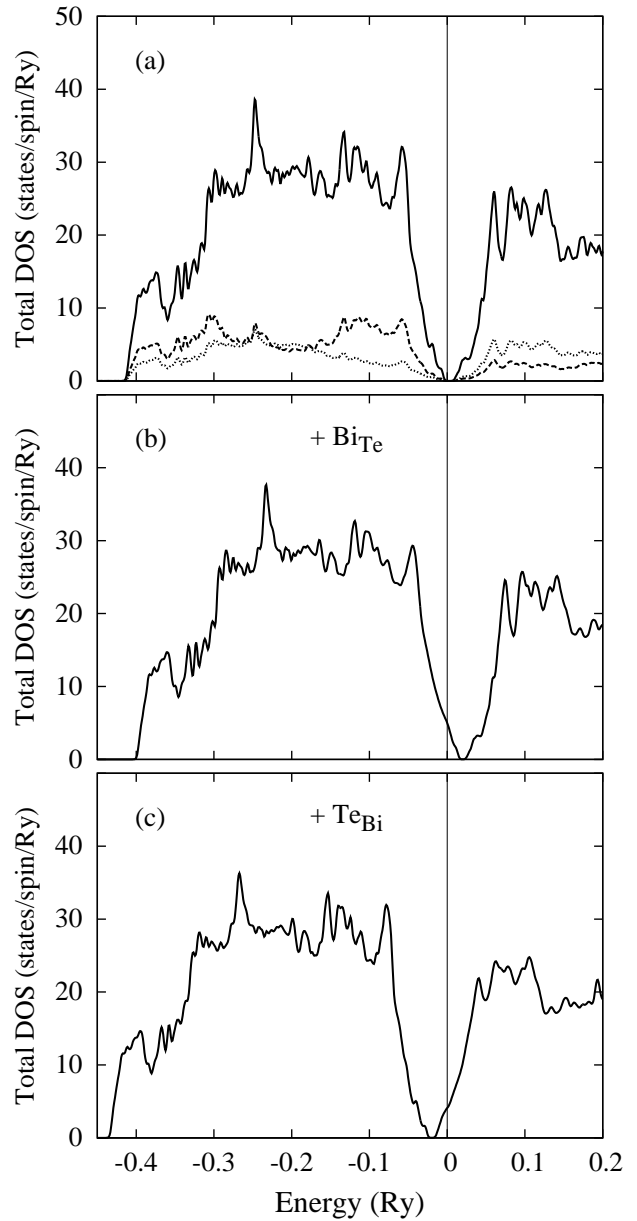}
\caption{Model A: Densities of states of doped Bi$_{2}$Te$_{3}$:
(a) Ideal Bi$_{2}$Te$_{3}$; (b) Bi$_{2}$Te$_{3}$ doped 
by \BiTe{} acceptors with concentrations $x$=0.025 on each of the
three Te-sublattices; and (c) Bi$_{2}$Te$_{3}$ doped by 
\TeBi{} donors with concentrations $x$=0.035 on each of the two 
Bi-sublattices. Total densities (full lines) and local densities 
on Bi (dotted lines) and Te (dashed lines) are shown for case (a). 
The Fermi energy is at zero.
}
\label{modelA}
\end{figure}

We present in Fig.~\ref{modelA} the total DOS of Bi$_{2}$Te$_{3}$
without and with antisite \BiTe{} and \TeBi{} defects.
The following conclusions can be done:
(i) The ideal Bi$_{2}$Te$_{3}$ is an insulator with a narrow gap
of 0.11~eV, which compares well with the experimental value of
0.16~eV \cite{r_57_Harma_PhysProp_BiTe_SbTe,r_09_Hsieh_ObservDiracCone_BiTe_SbTe}. The shape of the DOS agrees well with corresponding
LMTO bandstructure calculations,\cite{r_97_Mishr_Jepsen_ES_BiTe_BiSe} as well as with calculations based 
on the full-potential linearized augmented-plane-wave method,\cite{r_02_Larson_ES_BiX_vsPhotoEm} in particular around the gap which is relevant for transport. The band inversion has been obtained similarly to other calculations\cite{r_10_Zhang_FPstudy_BiTe_BiSe_SbTe_NJOP,r_13_Aguilera_Blugel_GW_TI_Bi2X3}. Decomposition of
the total DOS into the local Bi- (and Te-) DOS's reveals a dominating
Bi-$6p$ (Te-$5p$) character of the conduction (valence) bands;
(ii) \BiTe{}-antisites act as acceptors forming the \tcg{hole-type}
conductor while \TeBi{}-antisites act as donors and form the \tcg{electron-type}
conductor; and
(iii) The disorder in broad $p$-bands is generally weak and
the gap survives. One should expect relatively low resistivities
controlled in both bands by mobile $p$-orbital type carriers.

\begin{figure*}
\includegraphics[width=18cm] {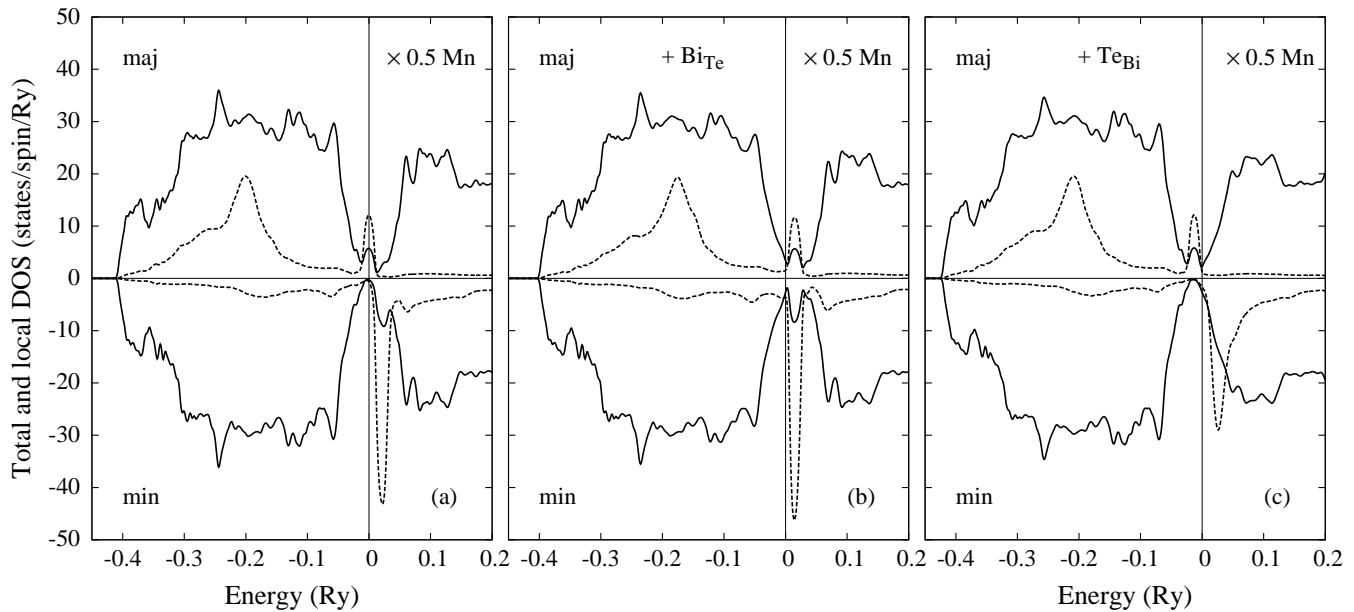}
\caption{Model B: (a) Spin-resolved densities of states of 
Bi$_{2}$Te$_{3}$ doped with \MnBi{} acceptors with concentration 
$x$=0.025 on each of two Bi-sublattices; (b) The same as in (a),
but co-doped with \BiTe{} acceptors with concentrations $x$=0.025 
on each of the three Te-sublattices; and (c) The same as in (a), but
co-doped with \TeBi{} donors with concentrations $x$=0.035 on
each of the two Bi-sublattices. Total densities (full lines) and 
local densities on Mn atoms (dashed lines, scaled by a factor 
0.5) are shown. The Fermi energy is at zero.
}
\label{modelB}
\end{figure*}

\subsubsection{Model B: Bi$_{2}$Te$_{3}$ with \MnBi{} defects}
\label{modB}

The DOS's for Bi$_{2}$Te$_{3}$ with substitutional Mn atoms on Bi-sites
without and with Bi/Te co-doping are shown in Fig.~\ref{modelB}.
We make the following comments:
(i) The Mn-doped crystal has halfmetallic character as the gap in the
minority states survives (Fig.~\ref{modelB}a);
(ii) A remarkable feature is the formation of the virtual bound state
at $E_{\rm F}$ in the majority states, while $E_{\rm F}$ lies in the
gap of the minority bands. The resistivity is thus only due to carriers
in the narrow Mn impurity band or, equivalently, due to low mobile 
$d$-electrons and it is thus expected to be large;
(iii) Co-doping by \BiTe{} acceptors (Fig.~\ref{modelB}b) shifts
$E_{\rm F}$ downwards from the impurity peak deeper into the
valence band so that carriers are partly mobile $p$-holes while
the majority and minority impurity peaks remain unoccupied. The 
downward shift of the Mn-minority peak kills the halfmetallic character 
of the alloy;
(iv) Co-doping with \TeBi{} donors (Fig.~\ref{modelB}c) acts in the
opposite way, namely, $E_{\rm F}$ is shifted upwards from the
majority impurity band into the conduction band, i.e., the
conductivity is due to mobile $p$-electrons. The alloy is nearly in a halfmetallic regime due to the
minority band gap located just below $E_{\rm F}$. In both doping models
one thus expects a reduction of the resistivity due to the reduced
impurity scatterings and mobile $p$-carriers. Of course, the value of
the resistivity in this case also depends on the amount of carriers
which is due to corresponding Mn and antisite concentrations;
and
\tcb{ (v) The neglect of lattice relaxations around \MnBi{} has a dramatic
effect on the position of Mn-impurity peaks with respect to
$E_{\rm F}$.\cite{r_Suppl-MnBiTe} }

\subsubsection{Model C: Bi$_{2}$Te$_{3}$ with Mn-interstitials}
\label{modC}

\begin{figure*}
\includegraphics[width=18cm] {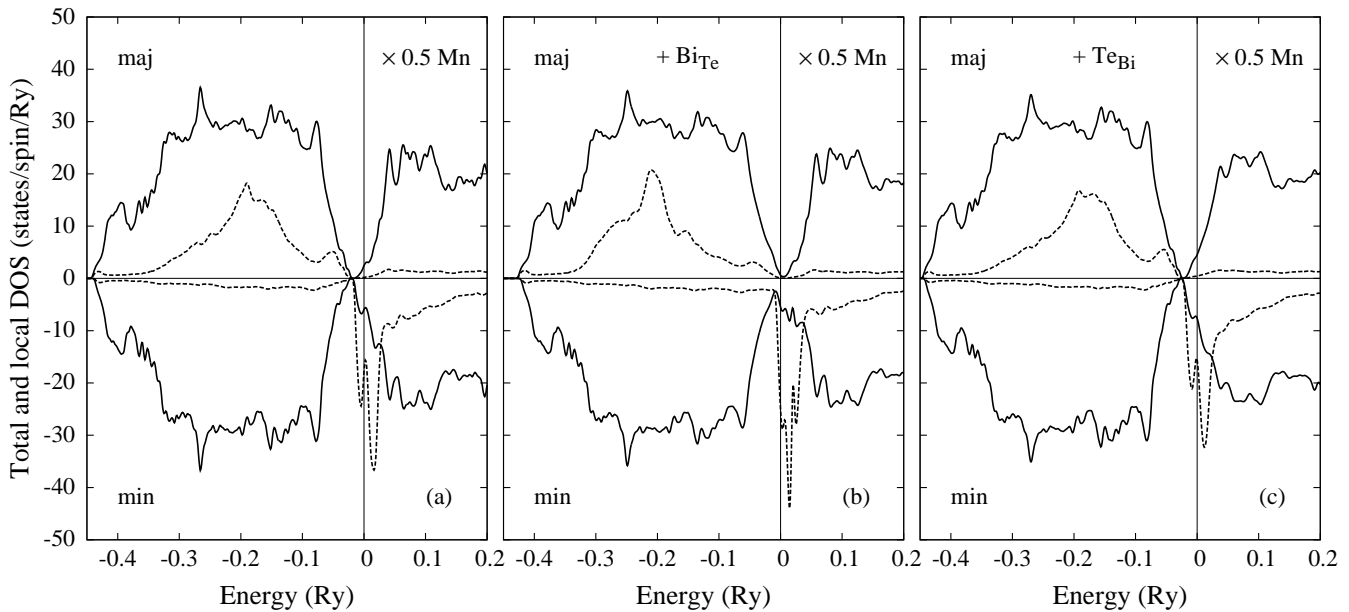}
\caption{Model C: (a) Spin-resolved densities of states of
Bi$_{2}$Te$_{3}$ doped with \MnI{} donors 
(van der Waals gap interstitial position) with concentration $x$=0.05; 
(b) The same as in (a), but co-doped with \BiTe{} acceptors 
with concentrations $x$=0.025 on each of the three Te-sublattices; 
and (c) The same as in (a), but co-doped with \TeBi{} donors 
with concentrations $x$=0.035 on each of the two Bi-sublattices. 
Total densities (full lines) and local densities on Mn atoms 
(dashed lines, scaled by a factor 0.5) are shown. The Fermi 
energy is at zero.
}
\label{modelC}
\end{figure*}

In Fig.~\ref{modelC} we show the DOS's for \MnI{} in
vdW gap positions without and with native defects.
The following conclusions are made:
(i) Donors shift $E_{\rm F}$ into the conduction band. Contrary
to the \MnBi{} doping we have now carriers in both majority and
minority bands. In this case the conductivity is partly due
the mobile $p$-electrons and should be \tcg{larger} than for the
Model B (without interstitials). In addition, the minority Mn-peak is quite broad
here. Note that the gap in the majority states survives; 
(ii) Co-doping by \BiTe{} acceptors shifts $E_{\rm F}$ into the 
majority gap while it lies partly in the impurity minority band, 
a situation similar to that in Fig.~\ref{modelB}a. In addition, 
the number of carriers (electrons in Fig.~\ref{modelC}a) is reduced 
by \BiTe{} acceptors which leads to a large sample resistivity;
(iii) On the contrary, \TeBi{} atoms donate additional carriers
(electrons) into the conduction band, but otherwise the situation is
similar to that in Fig.~\ref{modelC}a so that one should expect
further reduction of resistivity. Interestingly, gaps in both majority
and minority bands survive so that alloy is a pure \tcg{electron-type}
conductor; and
\tcb{ (iv) As mentioned above, there is a negligible effect of lattice
relaxations in this case. }

\subsubsection{Model D: Bi$_{2}$Te$_{3}$ with \MnBi{} defects
and Mn interstitials}
\label{modD}

\begin{figure}
\includegraphics[width=\columnwidth]{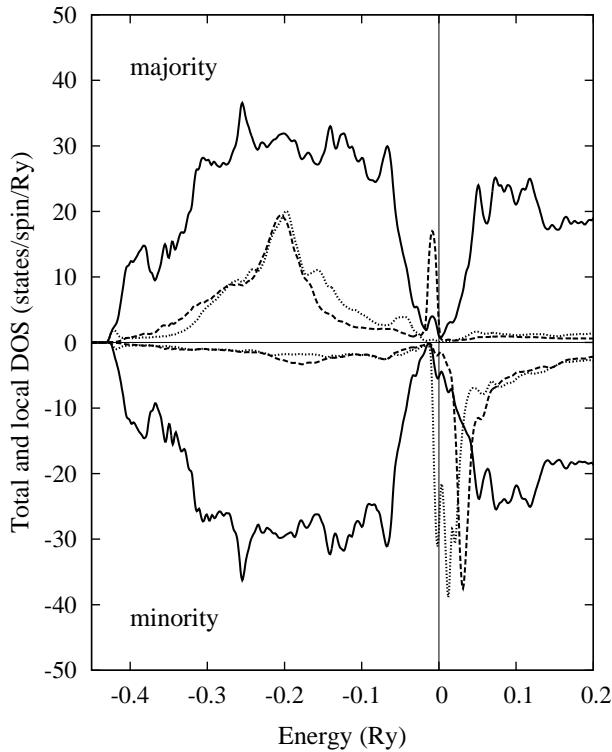}
\caption{ Model D: Spin-resolved densities of states of
Bi$_{2}$Te$_{3}$ doped by \MnBi{} acceptors with concentrations
$x$=0.0125 on each of the two Bi-sublattices and by  
\MnI{} donors with the concentration $x$=0.025. The total Mn
concentration is thus $x$=0.05 like in Models B and C. Total 
densities (full lines) and local densities on Mn atoms (dashed 
lines for \MnBi{} and dotted for \MnI{}, both scaled 
by a factor 0.5) are shown. The Fermi energy is at zero.
}
\label{modelD}
\end{figure}

Finally, we show in Fig.~\ref{modelD} a combined case when both
\MnBi{} and \MnI{} defects are present at the same
time, while preserving the same nominal Mn concentration as in the
previous cases ($x$=0.05). It should be noted that this can be 
achieved in many ways. Here we show the case when 
concentrations of Mn-substitutionals and Mn-interstitials are 
the same, namely, $x=0.025$ for each of these defects.
We make the following comments:
(i) The gap in the minority band survives, but $E_{\rm F}$ lies in the
conduction band, while the valence band including the virtual bound
state is fully occupied; and
(ii) The resistivity for small amounts of \MnI{} donors
should be large as $E_{\rm F}$ will be still partly in Mn $d$-impurity 
peak (low carrier mobility). The situation will be changed by 
increasing the amount of \MnI{} at the cost of \MnBi{}-defects. $E_{\rm F}$ will move away from impurity peak and there will be an increased amount of $p$-orbital carriers in the alloy. In other words, we can 
interpolate between the above two pure limits, Fig.~\ref{modelB}a and 
Fig.~\ref{modelC}a, respectively (see below corresponding
resistivities, Fig.~\ref{rho_si}).

\subsubsection{Magnetic moments}

We conclude this Section by showing in Table~\ref{tab_mm} the total magnetic moments per
Mn atom and formula unit, and the local magnetic moments on Mn atoms
for Models B, C, and D.
It should be noted that local moments depend on their definition, 
in particular on the space included in their calculations.
On the other hand the total moments estimated in different methods can
be compared reliably.
The total moments obtained in VASP calculations (for a slightly smaller Mn concentration $x$=0.042) for models B and C (without interstitials) are
3.97~$\mu_{\rm B}$ and 3.73~$\mu_{\rm B}$, respectively, in a good
agreement with the presented TB-LMTO values.
The co-doping by acceptors/donors in Model B shifts $E_{\rm F}$ down/up 
with respect to the undoped case such that it decreases/increases
the occupation of majority states, which explains the 
smaller/larger calculated moments, respectively.
The opposite trend is seen for Mn interstitials.
In the case of Model D a large occupation of majority states, which include
the impurity peak, leads to an increase of the total moment.

\begin{table}[tbp]
\caption{Total magnetic moments per Mn atom in formula unit 
($M_{\rm tot}$) and local magnetic moments on Mn atoms 
($m^{\rm Mn}$, in brackets) in $\mu_{\rm B}$ for Bi$_{2}$Te$_{3}$ 
crystal doped with Mn-atoms and native defects. Different models 
named Model B (with \MnBi{}), Model C (with \MnI{}), and Model D (with \MnBi{}, \MnI{}) and their types named (a), (b), and (c) are the same as those used in Figs.~\ref{modelB}, 
\ref{modelC}, and \ref{modelD}, respectively. There is only one 
type for the Model D denoted as (a) here, while corresponding local 
Mn moments in the bracket are those for \MnBi{}/\MnI{}, 
respectively. The concentrations $x$ of Mn dopants per formula 
unit are assumed to be $x$=0.05 for all models.
\\}
\renewcommand{\arraystretch}{1.2}
\begin{tabular}{|c|c|c|c|}
\hline
& \multicolumn{3}{|c|}{ Magnetic moments in Bi$_{2}$Te$_{3}$ } \\
\cline{2-4}
 Models/Types & (a) & (b) + \BiTe{} & (c) + \TeBi{} \\ \hline
 Model B  & 3.99 (3.36) & $\ $2.87 (2.77)$\ $ & $\ $4.81 (3.72)$\ $ \\
 Model C  & 3.57 (3.43) & 4.36 (3.63) & 3.30 (3.38) \\
 Model D  & 4.40 (3.70/3.56) &   $-$  &   $-$   \\ \hline
\end{tabular}
\label{tab_mm}
\end{table}

\begin{table}[tbp]
\caption{ Residual resistivities in the basal plane and total residual resistivities (in brackets) for Bi$_{2}$Te$_{3}$ crystal
(in m$\Omega$cm) doped with Mn-atoms and native defects. Different 
models named Model A, Model B (with \MnBi{}), Model C (with \MnI{}), and Model D (with \MnBi{}, \MnI{}) and their types 
named (a), (b), and (c) are the same as those used in 
Figs.~\ref{modelA}, \ref{modelB}, \ref{modelC}, and \ref{modelD}, 
respectively. There is only one type for the Model D denoted as 
(a) here. The same concentrations $x$ of Mn dopants per formula
unit are assumed in Models B, C, and D. 
\\}
\renewcommand{\arraystretch}{1.2}
\begin{tabular}{|c|c|c|c|}
\hline
& \multicolumn{3}{|c|}{ Resistivities in Bi$_{2}$Te$_{3}$ } \\
\cline{2-4}
 Models/Types & (a) & (b) + \BiTe{} & (c) + \TeBi{} \\ \hline
 Model A  & $ \infty$ ($ \infty$) &  $\ $0.13 (0.17)$\ $  &  $\ $0.17 (0.32)$\ $   \\
 Model B  &  $\ $2.45 (2.88)$\ $    &  0.49 (0.64)  &  0.38 (0.79)   \\
 Model C  &   0.11 (0.23)    & 2.95 (3.70)  &  0.24 (0.40)   \\
 Model D  &  1.52 (2.46)    &   $-$  &   $-$   \\ \hline
\end{tabular}
\label{tab_rho}
\end{table}

\subsection{Transport properties of Mn-doped Bi$_{2}$Te$_{3}$}
\label{transp}

Total residual resistivities for the studied Models A, B, C, and
D are listed in Table~\ref{tab_rho} assuming the same total Mn
concentration $x$ in the formula unit, namely, $x$=0.05.  
\tcg{With the knowledge of the above shown DOS's} the results can be 
qualitatively understood in terms of varying carrier 
concentrations due to different doping types (acceptors or 
donors) and relative positions of the Mn impurity level with 
respect to $E_{\rm F}$ (low mobility $d$-electron carriers vs high-mobility 
$p$-electron carriers).
We refer the reader to the discussion above (see Section
\ref{elstr}), which of course allows to understand only the relative
values of resistivities, not their quantitative 
values (listed in Table~\ref{tab_rho}).

\begin{figure}
\includegraphics[width=\columnwidth]{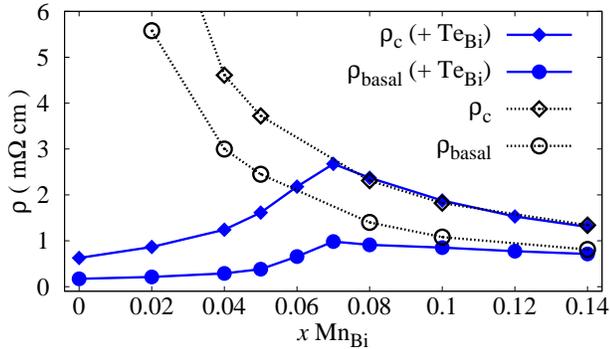}
\caption{ The concentration dependence of resistivities in the basal plane (circles) and along the $c$-axis (diamonds) in m$\Omega$cm for Bi$_{2}$Te$_{3}$ doped with \MnBi{} 
acceptors in a perfect Bi$_{2}$Te$_{3}$ crystal (dotted lines, empty symbols), and in the crystal with 3.5\% \TeBi{} antisites
(full lines, full symbols). The Mn concentration is per 
formula unit in both cases. 
}
\label{rho_s}
\end{figure}

\begin{figure}
\includegraphics[width=\columnwidth]{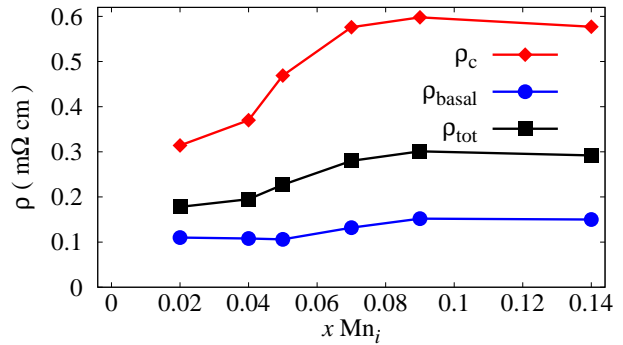}
\caption{ The concentration dependence of total resistivities
in m$\Omega$cm for Bi$_{2}$Te$_{3}$ doped with \MnI{} interstitials 
(crosses). The Mn concentration is per 
formula unit in both cases. Also shown are corresponding 
resistivities along the $c$-axis (full diamonds) and in 
the basal plane (full circles).
}
\label{rho_i}
\end{figure}

\begin{figure}
\includegraphics[width=\columnwidth]{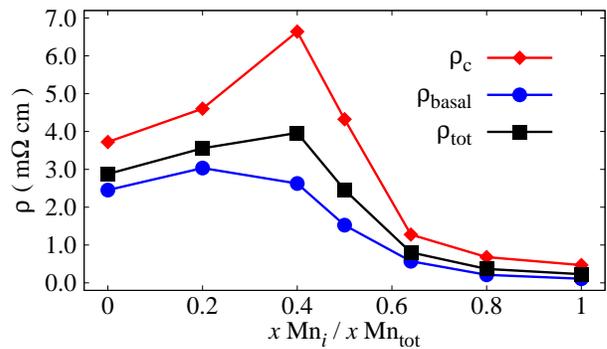}
\caption{ The total resistivities in m$\Omega$cm for 
Bi$_{2}$Te$_{3}$ as a function of the varying ratio between 
\MnI{} and \MnBi{} concentrations present in
the sample (crosses). The total Mn concentration $x$
is kept constant ($x$=0.05). Also shown are corresponding 
resistivities along the $c$-axis (full diamonds) and in 
the basal plane (full circles). 
}
\label{rho_si}
\end{figure}

For Models B and C, as well as for Model D, we have also
studied corresponding trends with changing Mn concentration.
Results are shown in Figs.~\ref{rho_s},~\ref{rho_i} and \ref{rho_si},
respectively.
A very different behavior for \MnBi{} substitutionals and \MnI{} interstitials
is obtained.
The resistivity of \MnBi{}-doped Bi$_{2}$Te$_{3}$ decreases monotonically
with Mn concentration similarly to the resistivity of Mn-doped GaAs 
alloys.\cite{r_04_tkd}
The resistivity is affected by the interplay of two factors, namely the increase of
strong impurity scattering at virtual bound states in the majority 
states and the increase of carrier (hole) concentration with a larger
Mn doping.
The latter effect dominates and the resistivity decreases with Mn
doping, exhibiting a non-metallic type of behavior.

\tcg{Notably the bulk conductivities measured experimentally for Bi$_{2}$Te$_{3}$  do not vanish
in undoped samples, because of a small amount of inevitable
native defects \cite{r_11_Hor_SC_SuppressBulkCond_Bi2X3,r_14_Hoefer_Cond_Via_SurfState_Bi2Te3}.
For this reason one also cannot expect a full agreement between a first
principles study assuming only Mn dopants and experiments for low Mn
concentration. 
The presence of native acceptors/donors shifts the Fermi level, can dramatically influence the resistivity (see Table~\ref{tab_rho}) and can be thus used as a tool to control it. We examine \MnBi{} doping in the presence of a fixed amount of \TeBi{} antisites, which leads to a behavior completely different from \MnBi{} doping in the ideal crystal. In this case the doping
only increases the amount of scattering, which provides a resistivity increase seen in Fig.~\ref{rho_s}. This is consistent with the experimentally observed dependencies on Mn doping \cite{r_10_Hor_Hasan_FM_TI_Mn-BiTe}. For high values of Mn doping (above $x$=0.05 in the studied case) the Fermi level is again in the impurity peak and the situation is comparable to \MnBi{} doping without \TeBi{} antisites. Note that for a wide range of
Mn content the antisite presence decreases the resistivity, contrary to the Mn-doped GaAs.\cite{r_04_tkd,r_07_ct_gamnas}  }

The resistivity of Bi$_{2}$Te$_{3}$ doped with \MnI{}
behaves differently from that of the \MnBi{} case, see Fig.~\ref{rho_i}.
First, the resistivity due to \MnBi{} doping is much larger as compared
to that due to \MnI{}, e.g., about one order of magnitude
for $x$=0.05.
This difference can be explained by the different character of the conductivity:
The \MnBi{} case corresponds to a halfmetal with $E_{\rm F}$ lying in the narrow virtual
bound state and the conductivity is dominated by low mobility $d$-holes.
In the case of \MnI{} interstitials both majority and minority channels
contribute, and the resistivity is much smaller due to mobile $p$-electrons.
When the dopant concentration increases, the carrier concentration increases,
but at the same time the impurity scattering at
Mn-minority impurity levels increases.
The competition of these two effects leads to an increase of the
resistivity for smaller Mn concentrations, but for higher ones there 
is a saturation and one can even observe a  small decrease at
large Mn doping.
On the other hand, absolute values of the resistivity remain much
smaller as compared to that due to \MnBi{} atoms.
A comparison with experiments is rather difficult, as reported values
vary for bulk samples and those prepared by molecular beam epitaxy.
Also, it is not clear what is the nature of native defects and
sometimes even the carrier type is changed during doping.
We have found in literature values ranging from 0.6~m$\Omega$cm \cite{r_14_Lee_Magnetism_transp_n-type_Mn-BiTe}
to 1.25~m$\Omega$cm \cite{r_10_Hor_Hasan_FM_TI_Mn-BiTe} or even to 2.5m$\Omega$cm.\cite{r_05_Choi_Mn-doped_V_VI_SpGlass}
Calculated values fall in this experimental range of resistivities.
A more detailed comparison is difficult due to insufficient details 
concerning types of defects, their types and amounts in the sample.

An alternative way is to vary the ratio between \MnBi{} and \MnI{} dopants
present together in the sample.
It should be noted that such ratios are not easy to control
intentionally as it is dictated by the thermodynamics of
the sample preparation and its annealing.
On the other hand, it is interesting theoretically to consider
this effect in some detail.
This is shown in Fig.~\ref{rho_si}, where we present calculated
resistivities for a fixed total Mn concentration $x$=0.05 as
a function of the ratio of concentrations of \MnI{}
($x_{\rm Mn_i}$) and Mn-substitutional ($x_{\rm Mn_{Bi}}$)
atoms.
We can thus interpolate between Model B and Model C.
An increased amount of donor-type Mn interstitials leads first
to the increase of the total resistivity due to the reduced
carrier concentration (\MnBi{} acceptor concentration is reduced
by donors), but a further increase of donors shifts $E_{\rm F}$
from the \MnBi{} virtual bound state to the conduction band
and both majority and minority channels now contribute to
the current, thus reducing the resistivity until we have a
low resistivity electron-like case with a high content of \MnI{} (see Fig.~\ref{modelC}a and Table~\ref{tab_rho}).
Note that the values of the resistivity vary by one order
of magnitude and reach the value of 1.25~m$\Omega$cm 
found experimentally.\cite{r_10_Hor_Hasan_FM_TI_Mn-BiTe}

\tcg{In  Figs.~\ref{rho_s}-\ref{rho_si} we also show the decomposition of the total 
resistivity into contributions along the $c$- and in the basal
plane  indicating a large value of the structural 
resistivity anisotropy. Notably the resistivity is always higher for the direction along the $c$-axis,
which is probably due to the smaller wavefunction overlap across the vdW gap.}

\tcg { Wei et al. \cite{r_15_Wei_TuningTranProp_Mn_Bi2Se3} have measured
samples of Mn-doped Bi$_{2}$Se$_{3}$ with both electron and hole
carriers, and obtained an order of magnitude higher resistivity for
hole-type conductor than for the electron-type one. Although the exact nature of defects
is not known there, this finding is in a qualitative agreement with our calculations
for Bi$_{2}$Te$_{3}$, where defects providing extra electrons
lead to an overall lower resistivity.  }

\section{Conclusions}
\label{concl}

We have demonstrated the ability of the TB-LMTO-CPA method to 
capture essential features of the electronic structure of
Bi$_{2}$Te$_{3}$ doped with Mn atoms.
An ideal Bi$_{2}$Te$_{3}$ is an insulator with a gap only 
slightly underestimated in comparison with the experiment.
We have reproduced the calculated magnetic moments and general DOS features for 
\MnBi{} acceptors and \MnI{} donors \tcg{employing analogous} full-potential supercell VASP 
calculations.
In particular, the \MnBi{} virtual bound state with Fermi energy
lying in the impurity peak \tcg{is obtained in both methods}, although in TB-LMTO-CPA
calculations the effect of the large mismatch of sizes 
of Bi- and Mn-atoms was included approximately.
This justifies the use of the TB-LMTO-CPA to estimate the 
electronic structure and for the first time transport properties also for 
more complex, but realistic cases: systems in which native antisite impurities (\BiTe{} donors and \TeBi{} acceptors) are present in addition to Mn atoms.

Antisite doping allows to control the number and the type of
carriers present in the sample and thus its resistivity.
This was documented by studying four different models, namely,
Bi$_{2}$Te$_{3}$ with/without native defects and \MnBi{} and
\MnI{} with/without native defects.
We have also considered alloys containing both types of Mn-dopants.
Detailed analysis of the DOS for the studied models allows to understand
relative values of corresponding resistivities depending on the 
character of their carriers, namely, low mobility $d$-electrons in 
impurity bands and high mobility $p$-holes/electrons in 
valence/conduction bands.
In general, low conductivities for \MnBi{} doping is due to the low 
mobility of dominating $d$-holes while much higher conductivity 
of \MnI{} is due to dominating $p$-electron transport. \tcg{Partial addition of \MnI{} into \MnBi{}-doped system leads to further decrease of conductivity as long as the Fermi level remains in the impurity peak.}
Overall \tcg{the variation of the amount of} native defects and/or partly also the amount of
\MnBi{} and \MnI{} in the sample allows to tune the
value of the conductivity \tcg{within a wide range of values. Another important feature is the high structural anisotropy of the conductivity tensor, with its components in the basal plane always higher than the one along the $c$-axis.}

\begin{acknowledgments}
This work was supported by the Czech Science Foundation Grant no.
14-30062S. JH acknowledges the Purkyn\v{e} fellowship program of the Czech Academy of Sciences.
\end{acknowledgments}

% \bibliographystyle{apsrev}
% \bibliography{b_kc}

\begin{thebibliography}{47}
\expandafter\ifx\csname natexlab\endcsname\relax\def\natexlab#1{#1}\fi
\expandafter\ifx\csname bibnamefont\endcsname\relax
  \def\bibnamefont#1{#1}\fi
\expandafter\ifx\csname bibfnamefont\endcsname\relax
  \def\bibfnamefont#1{#1}\fi
\expandafter\ifx\csname citenamefont\endcsname\relax
  \def\citenamefont#1{#1}\fi
\expandafter\ifx\csname url\endcsname\relax
  \def\url#1{\texttt{#1}}\fi
\expandafter\ifx\csname urlprefix\endcsname\relax\def\urlprefix{URL }\fi
\providecommand{\bibinfo}[2]{#2}
\providecommand{\eprint}[2][]{\url{#2}}

\bibitem[{\citenamefont{Dietl and Ohno}(2014)}]{r_14_Dietl_RMP_DilFMSC}
\bibinfo{author}{\bibfnamefont{T.}~\bibnamefont{Dietl}} \bibnamefont{and}
  \bibinfo{author}{\bibfnamefont{H.}~\bibnamefont{Ohno}},
  \bibinfo{journal}{Rev. Mod. Phys.} \textbf{\bibinfo{volume}{86}},
  \bibinfo{pages}{187} (\bibinfo{year}{2014}).

\bibitem[{\citenamefont{Sato et~al.}(2010)\citenamefont{Sato, Bergqvist,
  Kudrnovsk\'{y}, Dederichs, Eriksson, Turek, Sanyal, Bouzerar,
  Katayama-Yoshida, Dinh et~al.}}]{r_10_Sato_Kud_Turek_FP_DMS}
\bibinfo{author}{\bibfnamefont{K.}~\bibnamefont{Sato}},
  \bibinfo{author}{\bibfnamefont{L.}~\bibnamefont{Bergqvist}},
  \bibinfo{author}{\bibfnamefont{J.}~\bibnamefont{Kudrnovsk\'{y}}},
  \bibinfo{author}{\bibfnamefont{P.~H.} \bibnamefont{Dederichs}},
  \bibinfo{author}{\bibfnamefont{O.}~\bibnamefont{Eriksson}},
  \bibinfo{author}{\bibfnamefont{I.}~\bibnamefont{Turek}},
  \bibinfo{author}{\bibfnamefont{B.}~\bibnamefont{Sanyal}},
  \bibinfo{author}{\bibfnamefont{G.}~\bibnamefont{Bouzerar}},
  \bibinfo{author}{\bibfnamefont{H.}~\bibnamefont{Katayama-Yoshida}},
  \bibinfo{author}{\bibfnamefont{V.~A.} \bibnamefont{Dinh}},
  \bibnamefont{et~al.}, \bibinfo{journal}{Rev. Mod. Phys.}
  \textbf{\bibinfo{volume}{82}}, \bibinfo{pages}{1633} (\bibinfo{year}{2010}).

\bibitem[{\citenamefont{Noh et~al.}(2008)\citenamefont{Noh, Koh, Oh, Park, Kim,
  Rameau, Valla, Kidd, Johnson, Hu et~al.}}]{r_08_Noh_SOeff_Bi2Te3_ARPES}
\bibinfo{author}{\bibfnamefont{H.-J.} \bibnamefont{Noh}},
  \bibinfo{author}{\bibfnamefont{H.}~\bibnamefont{Koh}},
  \bibinfo{author}{\bibfnamefont{S.-J.} \bibnamefont{Oh}},
  \bibinfo{author}{\bibfnamefont{J.-H.} \bibnamefont{Park}},
  \bibinfo{author}{\bibfnamefont{H.-D.} \bibnamefont{Kim}},
  \bibinfo{author}{\bibfnamefont{J.~D.} \bibnamefont{Rameau}},
  \bibinfo{author}{\bibfnamefont{T.}~\bibnamefont{Valla}},
  \bibinfo{author}{\bibfnamefont{T.~E.} \bibnamefont{Kidd}},
  \bibinfo{author}{\bibfnamefont{P.~D.} \bibnamefont{Johnson}},
  \bibinfo{author}{\bibfnamefont{Y.}~\bibnamefont{Hu}}, \bibnamefont{et~al.},
  \bibinfo{journal}{EPL (Europhysics Letters)} \textbf{\bibinfo{volume}{81}},
  \bibinfo{pages}{57006} (\bibinfo{year}{2008}).

\bibitem[{\citenamefont{Fu et~al.}(2007)\citenamefont{Fu, Kane, and
  Mele}}]{r_07_Fu_Kane_3DTI}
\bibinfo{author}{\bibfnamefont{L.}~\bibnamefont{Fu}},
  \bibinfo{author}{\bibfnamefont{C.~L.} \bibnamefont{Kane}}, \bibnamefont{and}
  \bibinfo{author}{\bibfnamefont{E.~J.} \bibnamefont{Mele}},
  \bibinfo{journal}{Phys. Rev. Lett.} \textbf{\bibinfo{volume}{98}},
  \bibinfo{pages}{106803} (\bibinfo{year}{2007}).

\bibitem[{\citenamefont{Zhang et~al.}(2009)\citenamefont{Zhang, Liu, Qi, Dai,
  Fang, and Zhang}}]{r_09_Zhang_TI_BiTe_BiSe_SbTe_NPhys}
\bibinfo{author}{\bibfnamefont{H.}~\bibnamefont{Zhang}},
  \bibinfo{author}{\bibfnamefont{C.-X.} \bibnamefont{Liu}},
  \bibinfo{author}{\bibfnamefont{X.-L.} \bibnamefont{Qi}},
  \bibinfo{author}{\bibfnamefont{X.}~\bibnamefont{Dai}},
  \bibinfo{author}{\bibfnamefont{Z.}~\bibnamefont{Fang}}, \bibnamefont{and}
  \bibinfo{author}{\bibfnamefont{S.-C.} \bibnamefont{Zhang}},
  \bibinfo{journal}{Nat Phys} \textbf{\bibinfo{volume}{5}},
  \bibinfo{pages}{438} (\bibinfo{year}{2009}).

\bibitem[{\citenamefont{Hasan and Kane}(2010)}]{r_10_Hasan_TI_RevMod}
\bibinfo{author}{\bibfnamefont{M.~Z.} \bibnamefont{Hasan}} \bibnamefont{and}
  \bibinfo{author}{\bibfnamefont{C.~L.} \bibnamefont{Kane}},
  \bibinfo{journal}{Rev. Mod. Phys.} \textbf{\bibinfo{volume}{82}},
  \bibinfo{pages}{3045} (\bibinfo{year}{2010}).

\bibitem[{\citenamefont{Zhang et~al.}(2010)\citenamefont{Zhang, Yu, Zhang, Dai,
  and Fang}}]{r_10_Zhang_FPstudy_BiTe_BiSe_SbTe_NJOP}
\bibinfo{author}{\bibfnamefont{W.}~\bibnamefont{Zhang}},
  \bibinfo{author}{\bibfnamefont{R.}~\bibnamefont{Yu}},
  \bibinfo{author}{\bibfnamefont{H.-J.} \bibnamefont{Zhang}},
  \bibinfo{author}{\bibfnamefont{X.}~\bibnamefont{Dai}}, \bibnamefont{and}
  \bibinfo{author}{\bibfnamefont{Z.}~\bibnamefont{Fang}}, \bibinfo{journal}{New
  Journal of Physics} \textbf{\bibinfo{volume}{12}}, \bibinfo{pages}{065013}
  (\bibinfo{year}{2010}).

\bibitem[{\citenamefont{Hsieh et~al.}(2009)\citenamefont{Hsieh, Xia, Qian,
  Wray, Meier, Dil, Osterwalder, Patthey, Fedorov, Lin
  et~al.}}]{r_09_Hsieh_ObservDiracCone_BiTe_SbTe}
\bibinfo{author}{\bibfnamefont{D.}~\bibnamefont{Hsieh}},
  \bibinfo{author}{\bibfnamefont{Y.}~\bibnamefont{Xia}},
  \bibinfo{author}{\bibfnamefont{D.}~\bibnamefont{Qian}},
  \bibinfo{author}{\bibfnamefont{L.}~\bibnamefont{Wray}},
  \bibinfo{author}{\bibfnamefont{F.}~\bibnamefont{Meier}},
  \bibinfo{author}{\bibfnamefont{J.~H.} \bibnamefont{Dil}},
  \bibinfo{author}{\bibfnamefont{J.}~\bibnamefont{Osterwalder}},
  \bibinfo{author}{\bibfnamefont{L.}~\bibnamefont{Patthey}},
  \bibinfo{author}{\bibfnamefont{A.~V.} \bibnamefont{Fedorov}},
  \bibinfo{author}{\bibfnamefont{H.}~\bibnamefont{Lin}}, \bibnamefont{et~al.},
  \bibinfo{journal}{Phys. Rev. Lett.} \textbf{\bibinfo{volume}{103}},
  \bibinfo{pages}{146401} (\bibinfo{year}{2009}).

\bibitem[{\citenamefont{Hor et~al.}(2010)\citenamefont{Hor, Roushan,
  Beidenkopf, Seo, Qu, Checkelsky, Wray, Hsieh, Xia, Xu
  et~al.}}]{r_10_Hor_Hasan_FM_TI_Mn-BiTe}
\bibinfo{author}{\bibfnamefont{Y.~S.} \bibnamefont{Hor}},
  \bibinfo{author}{\bibfnamefont{P.}~\bibnamefont{Roushan}},
  \bibinfo{author}{\bibfnamefont{H.}~\bibnamefont{Beidenkopf}},
  \bibinfo{author}{\bibfnamefont{J.}~\bibnamefont{Seo}},
  \bibinfo{author}{\bibfnamefont{D.}~\bibnamefont{Qu}},
  \bibinfo{author}{\bibfnamefont{J.~G.} \bibnamefont{Checkelsky}},
  \bibinfo{author}{\bibfnamefont{L.~A.} \bibnamefont{Wray}},
  \bibinfo{author}{\bibfnamefont{D.}~\bibnamefont{Hsieh}},
  \bibinfo{author}{\bibfnamefont{Y.}~\bibnamefont{Xia}},
  \bibinfo{author}{\bibfnamefont{S.-Y.} \bibnamefont{Xu}},
  \bibnamefont{et~al.}, \bibinfo{journal}{Phys. Rev. B}
  \textbf{\bibinfo{volume}{81}}, \bibinfo{pages}{195203}
  (\bibinfo{year}{2010}).

\bibitem[{\citenamefont{Choi et~al.}(2005)\citenamefont{Choi, Lee, Kim, Choi,
  Choi, Song, and Cho}}]{r_05_Choi_Mn-doped_V_VI_SpGlass}
\bibinfo{author}{\bibfnamefont{J.}~\bibnamefont{Choi}},
  \bibinfo{author}{\bibfnamefont{H.-W.} \bibnamefont{Lee}},
  \bibinfo{author}{\bibfnamefont{B.-S.} \bibnamefont{Kim}},
  \bibinfo{author}{\bibfnamefont{S.}~\bibnamefont{Choi}},
  \bibinfo{author}{\bibfnamefont{J.}~\bibnamefont{Choi}},
  \bibinfo{author}{\bibfnamefont{J.~H.} \bibnamefont{Song}}, \bibnamefont{and}
  \bibinfo{author}{\bibfnamefont{S.}~\bibnamefont{Cho}},
  \bibinfo{journal}{Journal of Applied Physics} \textbf{\bibinfo{volume}{97}},
  \bibinfo{eid}{10D324} (\bibinfo{year}{2005}).

\bibitem[{\citenamefont{Lee et~al.}(2014)\citenamefont{Lee, Richardella, Rench,
  Fraleigh, Flanagan, Borchers, Tao, and
  Samarth}}]{r_14_Lee_Magnetism_transp_n-type_Mn-BiTe}
\bibinfo{author}{\bibfnamefont{J.~S.} \bibnamefont{Lee}},
  \bibinfo{author}{\bibfnamefont{A.}~\bibnamefont{Richardella}},
  \bibinfo{author}{\bibfnamefont{D.~W.} \bibnamefont{Rench}},
  \bibinfo{author}{\bibfnamefont{R.~D.} \bibnamefont{Fraleigh}},
  \bibinfo{author}{\bibfnamefont{T.~C.} \bibnamefont{Flanagan}},
  \bibinfo{author}{\bibfnamefont{J.~A.} \bibnamefont{Borchers}},
  \bibinfo{author}{\bibfnamefont{J.}~\bibnamefont{Tao}}, \bibnamefont{and}
  \bibinfo{author}{\bibfnamefont{N.}~\bibnamefont{Samarth}},
  \bibinfo{journal}{Phys. Rev. B} \textbf{\bibinfo{volume}{89}},
  \bibinfo{pages}{174425} (\bibinfo{year}{2014}).

\bibitem[{\citenamefont{Tarasenko et~al.}(2016)\citenamefont{Tarasenko,
  Vališka, Vondrá\v{c}ek, Horáková, Tká\v{c}, Carva, Baláž, Holý,
  Springholz, Sechovský et~al.}}]{r_16_Taras_cb_MagStruct_Mn_BiSe}
\bibinfo{author}{\bibfnamefont{R.}~\bibnamefont{Tarasenko}},
  \bibinfo{author}{\bibfnamefont{M.}~\bibnamefont{Vališka}},
  \bibinfo{author}{\bibfnamefont{M.}~\bibnamefont{Vondrá\v{c}ek}},
  \bibinfo{author}{\bibfnamefont{K.}~\bibnamefont{Horáková}},
  \bibinfo{author}{\bibfnamefont{V.}~\bibnamefont{Tká\v{c}}},
  \bibinfo{author}{\bibfnamefont{K.}~\bibnamefont{Carva}},
  \bibinfo{author}{\bibfnamefont{P.}~\bibnamefont{Baláž}},
  \bibinfo{author}{\bibfnamefont{V.}~\bibnamefont{Holý}},
  \bibinfo{author}{\bibfnamefont{G.}~\bibnamefont{Springholz}},
  \bibinfo{author}{\bibfnamefont{V.}~\bibnamefont{Sechovský}},
  \bibnamefont{et~al.}, \bibinfo{journal}{Physica B: Condensed Matter}
  \textbf{\bibinfo{volume}{481}}, \bibinfo{pages}{262} (\bibinfo{year}{2016}).

\bibitem[{\citenamefont{Vergniory et~al.}(2014)\citenamefont{Vergniory,
  Otrokov, Thonig, Hoffmann, Maznichenko, Geilhufe, Zubizarreta, Ostanin,
  Marmodoro, Henk et~al.}}]{r_14_Vergn_Mertig_ExchInt_BiSe_BiTe_SbTe}
\bibinfo{author}{\bibfnamefont{M.~G.} \bibnamefont{Vergniory}},
  \bibinfo{author}{\bibfnamefont{M.~M.} \bibnamefont{Otrokov}},
  \bibinfo{author}{\bibfnamefont{D.}~\bibnamefont{Thonig}},
  \bibinfo{author}{\bibfnamefont{M.}~\bibnamefont{Hoffmann}},
  \bibinfo{author}{\bibfnamefont{I.~V.} \bibnamefont{Maznichenko}},
  \bibinfo{author}{\bibfnamefont{M.}~\bibnamefont{Geilhufe}},
  \bibinfo{author}{\bibfnamefont{X.}~\bibnamefont{Zubizarreta}},
  \bibinfo{author}{\bibfnamefont{S.}~\bibnamefont{Ostanin}},
  \bibinfo{author}{\bibfnamefont{A.}~\bibnamefont{Marmodoro}},
  \bibinfo{author}{\bibfnamefont{J.}~\bibnamefont{Henk}}, \bibnamefont{et~al.},
  \bibinfo{journal}{Phys. Rev. B} \textbf{\bibinfo{volume}{89}},
  \bibinfo{pages}{165202} (\bibinfo{year}{2014}).

\bibitem[{\citenamefont{Henk et~al.}(2012)\citenamefont{Henk, Flieger,
  Maznichenko, Mertig, Ernst, Eremeev, and
  Chulkov}}]{r_12_Henk_TopChar_Dirac_Mn-BiTe}
\bibinfo{author}{\bibfnamefont{J.}~\bibnamefont{Henk}},
  \bibinfo{author}{\bibfnamefont{M.}~\bibnamefont{Flieger}},
  \bibinfo{author}{\bibfnamefont{I.~V.} \bibnamefont{Maznichenko}},
  \bibinfo{author}{\bibfnamefont{I.}~\bibnamefont{Mertig}},
  \bibinfo{author}{\bibfnamefont{A.}~\bibnamefont{Ernst}},
  \bibinfo{author}{\bibfnamefont{S.~V.} \bibnamefont{Eremeev}},
  \bibnamefont{and} \bibinfo{author}{\bibfnamefont{E.~V.}
  \bibnamefont{Chulkov}}, \bibinfo{journal}{Phys. Rev. Lett.}
  \textbf{\bibinfo{volume}{109}}, \bibinfo{pages}{076801}
  (\bibinfo{year}{2012}).

\bibitem[{\citenamefont{Watson et~al.}(2013)\citenamefont{Watson,
  Collins-McIntyre, Shelford, Coldea, Prabhakaran, Speller, Mousavi, Grovenor,
  Salman, Giblin et~al.}}]{r_13_Watso_BiTeMn-Prop-Istit}
\bibinfo{author}{\bibfnamefont{M.~D.} \bibnamefont{Watson}},
  \bibinfo{author}{\bibfnamefont{L.~J.} \bibnamefont{Collins-McIntyre}},
  \bibinfo{author}{\bibfnamefont{L.~R.} \bibnamefont{Shelford}},
  \bibinfo{author}{\bibfnamefont{A.}~\bibnamefont{Coldea}},
  \bibinfo{author}{\bibfnamefont{D.}~\bibnamefont{Prabhakaran}},
  \bibinfo{author}{\bibfnamefont{S.~C.} \bibnamefont{Speller}},
  \bibinfo{author}{\bibfnamefont{T.}~\bibnamefont{Mousavi}},
  \bibinfo{author}{\bibfnamefont{C.}~\bibnamefont{Grovenor}},
  \bibinfo{author}{\bibfnamefont{Z.}~\bibnamefont{Salman}},
  \bibinfo{author}{\bibfnamefont{S.~R.} \bibnamefont{Giblin}},
  \bibnamefont{et~al.}, \bibinfo{journal}{New Journal of Physics}
  \textbf{\bibinfo{volume}{15}}, \bibinfo{pages}{103016}
  (\bibinfo{year}{2013}).

\bibitem[{\citenamefont{R\r{u}\v{z}i\v{c}ka
  et~al.}(2015)\citenamefont{R\r{u}\v{z}i\v{c}ka, Caha, Hol\'{y}, Steiner,
  Volobuiev, Ney, Bauer, Ducho\v{n}, Veltrusk\'{a}, Khalakhan
  et~al.}}]{r_15_Ruzicka_Holy_Mn-BiTe_properties}
\bibinfo{author}{\bibfnamefont{J.}~\bibnamefont{R\r{u}\v{z}i\v{c}ka}},
  \bibinfo{author}{\bibfnamefont{O.}~\bibnamefont{Caha}},
  \bibinfo{author}{\bibfnamefont{V.}~\bibnamefont{Hol\'{y}}},
  \bibinfo{author}{\bibfnamefont{H.}~\bibnamefont{Steiner}},
  \bibinfo{author}{\bibfnamefont{V.}~\bibnamefont{Volobuiev}},
  \bibinfo{author}{\bibfnamefont{A.}~\bibnamefont{Ney}},
  \bibinfo{author}{\bibfnamefont{G.}~\bibnamefont{Bauer}},
  \bibinfo{author}{\bibfnamefont{T.}~\bibnamefont{Ducho\v{n}}},
  \bibinfo{author}{\bibfnamefont{K.}~\bibnamefont{Veltrusk\'{a}}},
  \bibinfo{author}{\bibfnamefont{I.}~\bibnamefont{Khalakhan}},
  \bibnamefont{et~al.}, \bibinfo{journal}{New Journal of Physics}
  \textbf{\bibinfo{volume}{17}}, \bibinfo{pages}{013028}
  (\bibinfo{year}{2015}).

\bibitem[{\citenamefont{Zhang et~al.}(2013)\citenamefont{Zhang, Ming, Huang,
  Liu, Kou, Fan, Wang, and Yao}}]{r_13_Zhang_MagDop_BiSe_BiTe_SbTe_FormEn}
\bibinfo{author}{\bibfnamefont{J.-M.} \bibnamefont{Zhang}},
  \bibinfo{author}{\bibfnamefont{W.}~\bibnamefont{Ming}},
  \bibinfo{author}{\bibfnamefont{Z.}~\bibnamefont{Huang}},
  \bibinfo{author}{\bibfnamefont{G.-B.} \bibnamefont{Liu}},
  \bibinfo{author}{\bibfnamefont{X.}~\bibnamefont{Kou}},
  \bibinfo{author}{\bibfnamefont{Y.}~\bibnamefont{Fan}},
  \bibinfo{author}{\bibfnamefont{K.~L.} \bibnamefont{Wang}}, \bibnamefont{and}
  \bibinfo{author}{\bibfnamefont{Y.}~\bibnamefont{Yao}},
  \bibinfo{journal}{Phys. Rev. B} \textbf{\bibinfo{volume}{88}},
  \bibinfo{pages}{235131} (\bibinfo{year}{2013}).

\bibitem[{\citenamefont{Scanlon et~al.}(2012)\citenamefont{Scanlon, King,
  Singh, de~la Torre, Walker, Balakrishnan, Baumberger, and
  Catlow}}]{r_12_Scanlon_TIBulkCon_Antisite}
\bibinfo{author}{\bibfnamefont{D.~O.} \bibnamefont{Scanlon}},
  \bibinfo{author}{\bibfnamefont{P.~D.~C.} \bibnamefont{King}},
  \bibinfo{author}{\bibfnamefont{R.~P.} \bibnamefont{Singh}},
  \bibinfo{author}{\bibfnamefont{A.}~\bibnamefont{de~la Torre}},
  \bibinfo{author}{\bibfnamefont{S.~M.} \bibnamefont{Walker}},
  \bibinfo{author}{\bibfnamefont{G.}~\bibnamefont{Balakrishnan}},
  \bibinfo{author}{\bibfnamefont{F.}~\bibnamefont{Baumberger}},
  \bibnamefont{and} \bibinfo{author}{\bibfnamefont{C.~R.~A.}
  \bibnamefont{Catlow}}, \bibinfo{journal}{Advanced Materials}
  \textbf{\bibinfo{volume}{24}}, \bibinfo{pages}{2154} (\bibinfo{year}{2012}).

\bibitem[{\citenamefont{Hor et~al.}(2011)\citenamefont{Hor, Checkelsky, Qu,
  Ong, and Cava}}]{r_11_Hor_SC_SuppressBulkCond_Bi2X3}
\bibinfo{author}{\bibfnamefont{Y.}~\bibnamefont{Hor}},
  \bibinfo{author}{\bibfnamefont{J.}~\bibnamefont{Checkelsky}},
  \bibinfo{author}{\bibfnamefont{D.}~\bibnamefont{Qu}},
  \bibinfo{author}{\bibfnamefont{N.}~\bibnamefont{Ong}}, \bibnamefont{and}
  \bibinfo{author}{\bibfnamefont{R.}~\bibnamefont{Cava}},
  \bibinfo{journal}{Journal of Physics and Chemistry of Solids}
  \textbf{\bibinfo{volume}{72}}, \bibinfo{pages}{572} (\bibinfo{year}{2011}).

\bibitem[{\citenamefont{Hoefer et~al.}(2014)\citenamefont{Hoefer, Becker, Rata,
  Swanson, Thalmeier, and Tjeng}}]{r_14_Hoefer_Cond_Via_SurfState_Bi2Te3}
\bibinfo{author}{\bibfnamefont{K.}~\bibnamefont{Hoefer}},
  \bibinfo{author}{\bibfnamefont{C.}~\bibnamefont{Becker}},
  \bibinfo{author}{\bibfnamefont{D.}~\bibnamefont{Rata}},
  \bibinfo{author}{\bibfnamefont{J.}~\bibnamefont{Swanson}},
  \bibinfo{author}{\bibfnamefont{P.}~\bibnamefont{Thalmeier}},
  \bibnamefont{and} \bibinfo{author}{\bibfnamefont{L.~H.} \bibnamefont{Tjeng}},
  \bibinfo{journal}{Proceedings of the National Academy of Sciences}
  \textbf{\bibinfo{volume}{111}}, \bibinfo{pages}{14979}
  (\bibinfo{year}{2014}).

\bibitem[{\citenamefont{Brahlek et~al.}(2015)\citenamefont{Brahlek, Koirala,
  Bansal, and Oh}}]{r_15_Brahl_TransportTI_Met2Ins}
\bibinfo{author}{\bibfnamefont{M.}~\bibnamefont{Brahlek}},
  \bibinfo{author}{\bibfnamefont{N.}~\bibnamefont{Koirala}},
  \bibinfo{author}{\bibfnamefont{N.}~\bibnamefont{Bansal}}, \bibnamefont{and}
  \bibinfo{author}{\bibfnamefont{S.}~\bibnamefont{Oh}}, \bibinfo{journal}{Solid
  State Communications} \textbf{\bibinfo{volume}{215-216}}, \bibinfo{pages}{54}
  (\bibinfo{year}{2015}).

\bibitem[{\citenamefont{Oh et~al.}(2014)\citenamefont{Oh, Son, Kim, Park, Min,
  and Lee}}]{r_14_Oh_Lee_Antisite_BiTe_BiSe}
\bibinfo{author}{\bibfnamefont{M.~W.} \bibnamefont{Oh}},
  \bibinfo{author}{\bibfnamefont{J.~H.} \bibnamefont{Son}},
  \bibinfo{author}{\bibfnamefont{B.~S.} \bibnamefont{Kim}},
  \bibinfo{author}{\bibfnamefont{S.~D.} \bibnamefont{Park}},
  \bibinfo{author}{\bibfnamefont{B.~K.} \bibnamefont{Min}}, \bibnamefont{and}
  \bibinfo{author}{\bibfnamefont{H.~W.} \bibnamefont{Lee}},
  \bibinfo{journal}{Journal of Applied Physics} \textbf{\bibinfo{volume}{115}},
  \bibinfo{eid}{133706} (\bibinfo{year}{2014}).

\bibitem[{\citenamefont{Soven}(1967)}]{r_67_ps}
\bibinfo{author}{\bibfnamefont{P.}~\bibnamefont{Soven}},
  \bibinfo{journal}{Phys.\ Rev.} \textbf{\bibinfo{volume}{156}},
  \bibinfo{pages}{809} (\bibinfo{year}{1967}).

\bibitem[{\citenamefont{Skriver}(1984)}]{r_84_Skriver_LMTO_book}
\bibinfo{author}{\bibfnamefont{H.}~\bibnamefont{Skriver}},
  \emph{\bibinfo{title}{The LMTO Method: Muffin-Tin Orbitals and Electronic
  Structure}} (\bibinfo{publisher}{Springer-Verlag, Berlin},
  \bibinfo{year}{1984}).

\bibitem[{\citenamefont{Turek et~al.}(1997)\citenamefont{Turek, Drchal,
  Kudrnovsk\'y, \v{S}ob, and Weinberger}}]{r_97_tdk}
\bibinfo{author}{\bibfnamefont{I.}~\bibnamefont{Turek}},
  \bibinfo{author}{\bibfnamefont{V.}~\bibnamefont{Drchal}},
  \bibinfo{author}{\bibfnamefont{J.}~\bibnamefont{Kudrnovsk\'y}},
  \bibinfo{author}{\bibfnamefont{M.}~\bibnamefont{\v{S}ob}}, \bibnamefont{and}
  \bibinfo{author}{\bibfnamefont{P.}~\bibnamefont{Weinberger}},
  \emph{\bibinfo{title}{Electronic Structure of Disordered Alloys, Surfaces and
  Interfaces}} (\bibinfo{publisher}{Kluwer, Boston}, \bibinfo{year}{1997}).

\bibitem[{\citenamefont{Vosko et~al.}(1980)\citenamefont{Vosko, Wilk, and
  Nusair}}]{r_80_vwn}
\bibinfo{author}{\bibfnamefont{S.~H.} \bibnamefont{Vosko}},
  \bibinfo{author}{\bibfnamefont{L.}~\bibnamefont{Wilk}}, \bibnamefont{and}
  \bibinfo{author}{\bibfnamefont{M.}~\bibnamefont{Nusair}},
  \bibinfo{journal}{Can.\ J. Phys.} \textbf{\bibinfo{volume}{58}},
  \bibinfo{pages}{1200} (\bibinfo{year}{1980}).

\bibitem[{\citenamefont{Turek et~al.}(2008)\citenamefont{Turek, Drchal, and
  Kudrnovsk\'{y}}}]{r_08_tdk_LMTO_SO}
\bibinfo{author}{\bibfnamefont{I.}~\bibnamefont{Turek}},
  \bibinfo{author}{\bibfnamefont{V.}~\bibnamefont{Drchal}}, \bibnamefont{and}
  \bibinfo{author}{\bibfnamefont{J.}~\bibnamefont{Kudrnovsk\'{y}}},
  \bibinfo{journal}{Philosophical Magazine} \textbf{\bibinfo{volume}{88}},
  \bibinfo{pages}{2787} (\bibinfo{year}{2008}).

\bibitem[{\citenamefont{Korzhavyi et~al.}(1995)\citenamefont{Korzhavyi, Ruban,
  Abrikosov, and Skriver}}]{r_95_Korzh_Ruban_ScrImp_Madelung_CPA}
\bibinfo{author}{\bibfnamefont{P.~A.} \bibnamefont{Korzhavyi}},
  \bibinfo{author}{\bibfnamefont{A.~V.} \bibnamefont{Ruban}},
  \bibinfo{author}{\bibfnamefont{I.~A.} \bibnamefont{Abrikosov}},
  \bibnamefont{and} \bibinfo{author}{\bibfnamefont{H.~L.}
  \bibnamefont{Skriver}}, \bibinfo{journal}{Phys. Rev. B}
  \textbf{\bibinfo{volume}{51}}, \bibinfo{pages}{5773} (\bibinfo{year}{1995}).

\bibitem[{\citenamefont{Kresse and Joubert}(1999)}]{r_99_Kresse_PAW}
\bibinfo{author}{\bibfnamefont{G.}~\bibnamefont{Kresse}} \bibnamefont{and}
  \bibinfo{author}{\bibfnamefont{D.}~\bibnamefont{Joubert}},
  \bibinfo{journal}{Phys. Rev. B} \textbf{\bibinfo{volume}{59}},
  \bibinfo{pages}{1758} (\bibinfo{year}{1999}).

\bibitem[{\citenamefont{Perdew and Zunger}(1981)}]{r_81_Perdew_Zunger_SIC_DFT}
\bibinfo{author}{\bibfnamefont{J.~P.} \bibnamefont{Perdew}} \bibnamefont{and}
  \bibinfo{author}{\bibfnamefont{A.}~\bibnamefont{Zunger}},
  \bibinfo{journal}{Phys. Rev. B} \textbf{\bibinfo{volume}{23}},
  \bibinfo{pages}{5048} (\bibinfo{year}{1981}).

\bibitem[{\citenamefont{Ceperley and Alder}(1980)}]{r_80_Ceperley_Alder_LDA_GS}
\bibinfo{author}{\bibfnamefont{D.~M.} \bibnamefont{Ceperley}} \bibnamefont{and}
  \bibinfo{author}{\bibfnamefont{B.~J.} \bibnamefont{Alder}},
  \bibinfo{journal}{Phys. Rev. Lett.} \textbf{\bibinfo{volume}{45}},
  \bibinfo{pages}{566} (\bibinfo{year}{1980}).

\bibitem[{\citenamefont{Perdew et~al.}(1996)\citenamefont{Perdew, Burke, and
  Ernzerhof}}]{r_96_Perdew_GGAsimple}
\bibinfo{author}{\bibfnamefont{J.~P.} \bibnamefont{Perdew}},
  \bibinfo{author}{\bibfnamefont{K.}~\bibnamefont{Burke}}, \bibnamefont{and}
  \bibinfo{author}{\bibfnamefont{M.}~\bibnamefont{Ernzerhof}},
  \bibinfo{journal}{Phys. Rev. Lett.} \textbf{\bibinfo{volume}{77}},
  \bibinfo{pages}{3865} (\bibinfo{year}{1996}).

\bibitem[{\citenamefont{Li et~al.}(2014)\citenamefont{Li, Zou, Li, and
  Zhou}}]{r_14_Li_FM_SurfS_Mn-BiTe_DFT}
\bibinfo{author}{\bibfnamefont{Y.}~\bibnamefont{Li}},
  \bibinfo{author}{\bibfnamefont{X.}~\bibnamefont{Zou}},
  \bibinfo{author}{\bibfnamefont{J.}~\bibnamefont{Li}}, \bibnamefont{and}
  \bibinfo{author}{\bibfnamefont{G.}~\bibnamefont{Zhou}}, \bibinfo{journal}{The
  Journal of Chemical Physics} \textbf{\bibinfo{volume}{140}},
  \bibinfo{eid}{124704} (\bibinfo{year}{2014}).

\bibitem[{\citenamefont{Kudrnovsk\'y and
  Drchal}(1990)}]{r_90_kd_LMTO_CPA_diffWS}
\bibinfo{author}{\bibfnamefont{J.}~\bibnamefont{Kudrnovsk\'y}}
  \bibnamefont{and} \bibinfo{author}{\bibfnamefont{V.}~\bibnamefont{Drchal}},
  \bibinfo{journal}{Phys. Rev. B} \textbf{\bibinfo{volume}{41}},
  \bibinfo{pages}{7515} (\bibinfo{year}{1990}).

\bibitem[{\citenamefont{Kudrnovsk\'y et~al.}(2009)\citenamefont{Kudrnovsk\'y,
  M\'aca, Turek, and Redinger}}]{r_09_kmt_Redin_AFM_Fe_Ir001}
\bibinfo{author}{\bibfnamefont{J.}~\bibnamefont{Kudrnovsk\'y}},
  \bibinfo{author}{\bibfnamefont{F.}~\bibnamefont{M\'aca}},
  \bibinfo{author}{\bibfnamefont{I.}~\bibnamefont{Turek}}, \bibnamefont{and}
  \bibinfo{author}{\bibfnamefont{J.}~\bibnamefont{Redinger}},
  \bibinfo{journal}{Phys. Rev. B} \textbf{\bibinfo{volume}{80}},
  \bibinfo{pages}{064405} (\bibinfo{year}{2009}).

\bibitem[{\citenamefont{Persson and
  Zunger}(2003)}]{r_03_Perss_Zunger_N_states_GaAsN}
\bibinfo{author}{\bibfnamefont{C.}~\bibnamefont{Persson}} \bibnamefont{and}
  \bibinfo{author}{\bibfnamefont{A.}~\bibnamefont{Zunger}},
  \bibinfo{journal}{Phys. Rev. B} \textbf{\bibinfo{volume}{68}},
  \bibinfo{pages}{035212} (\bibinfo{year}{2003}).

\bibitem[{r_S()}]{r_Suppl-MnBiTe}
\bibinfo{note}{See Supplemental Material at [URL will be inserted by publisher]
  for details of the treatment of alloy disorder with unequal constituent atom
  radii}.

\bibitem[{\citenamefont{Turek and
  Zalezak}(2010)}]{r_10_t_Zalezak_ReRe_CoNi_CuNi}
\bibinfo{author}{\bibfnamefont{I.}~\bibnamefont{Turek}} \bibnamefont{and}
  \bibinfo{author}{\bibfnamefont{T.}~\bibnamefont{Zalezak}},
  \bibinfo{journal}{Journal of Physics: Conference Series}
  \textbf{\bibinfo{volume}{200}}, \bibinfo{pages}{052029}
  (\bibinfo{year}{2010}).

\bibitem[{\citenamefont{Turek et~al.}(2002)\citenamefont{Turek, Kudrnovsk\'y,
  Drchal, Szunyogh, and Weinberger}}]{r_02_tkd}
\bibinfo{author}{\bibfnamefont{I.}~\bibnamefont{Turek}},
  \bibinfo{author}{\bibfnamefont{J.}~\bibnamefont{Kudrnovsk\'y}},
  \bibinfo{author}{\bibfnamefont{V.}~\bibnamefont{Drchal}},
  \bibinfo{author}{\bibfnamefont{L.}~\bibnamefont{Szunyogh}}, \bibnamefont{and}
  \bibinfo{author}{\bibfnamefont{P.}~\bibnamefont{Weinberger}},
  \bibinfo{journal}{Phys.\ Rev.\ B} \textbf{\bibinfo{volume}{65}},
  \bibinfo{pages}{125101} (\bibinfo{year}{2002}).

\bibitem[{\citenamefont{Carva et~al.}(2006)\citenamefont{Carva, Turek,
  Kudrnovsk\'y, and Bengone}}]{r_06_ctkb_vertex}
\bibinfo{author}{\bibfnamefont{K.}~\bibnamefont{Carva}},
  \bibinfo{author}{\bibfnamefont{I.}~\bibnamefont{Turek}},
  \bibinfo{author}{\bibfnamefont{J.}~\bibnamefont{Kudrnovsk\'y}},
  \bibnamefont{and} \bibinfo{author}{\bibfnamefont{O.}~\bibnamefont{Bengone}},
  \bibinfo{journal}{Phys.\ Rev.\ B} \textbf{\bibinfo{volume}{73}},
  \bibinfo{pages}{144421} (\bibinfo{year}{2006}).

\bibitem[{\citenamefont{Harman et~al.}(1957)\citenamefont{Harman, Paris,
  Miller, and Goering}}]{r_57_Harma_PhysProp_BiTe_SbTe}
\bibinfo{author}{\bibfnamefont{T.}~\bibnamefont{Harman}},
  \bibinfo{author}{\bibfnamefont{B.}~\bibnamefont{Paris}},
  \bibinfo{author}{\bibfnamefont{S.}~\bibnamefont{Miller}}, \bibnamefont{and}
  \bibinfo{author}{\bibfnamefont{H.}~\bibnamefont{Goering}},
  \bibinfo{journal}{Journal of Physics and Chemistry of Solids}
  \textbf{\bibinfo{volume}{2}}, \bibinfo{pages}{181 } (\bibinfo{year}{1957}).

\bibitem[{\citenamefont{Mishra et~al.}(1997)\citenamefont{Mishra, Satpathy, and
  Jepsen}}]{r_97_Mishr_Jepsen_ES_BiTe_BiSe}
\bibinfo{author}{\bibfnamefont{S.~K.} \bibnamefont{Mishra}},
  \bibinfo{author}{\bibfnamefont{S.}~\bibnamefont{Satpathy}}, \bibnamefont{and}
  \bibinfo{author}{\bibfnamefont{O.}~\bibnamefont{Jepsen}},
  \bibinfo{journal}{Journal of Physics: Condensed Matter}
  \textbf{\bibinfo{volume}{9}}, \bibinfo{pages}{461} (\bibinfo{year}{1997}).

\bibitem[{\citenamefont{Larson et~al.}(2002)\citenamefont{Larson, Greanya,
  Tonjes, Liu, Mahanti, and Olson}}]{r_02_Larson_ES_BiX_vsPhotoEm}
\bibinfo{author}{\bibfnamefont{P.}~\bibnamefont{Larson}},
  \bibinfo{author}{\bibfnamefont{V.~A.} \bibnamefont{Greanya}},
  \bibinfo{author}{\bibfnamefont{W.~C.} \bibnamefont{Tonjes}},
  \bibinfo{author}{\bibfnamefont{R.}~\bibnamefont{Liu}},
  \bibinfo{author}{\bibfnamefont{S.~D.} \bibnamefont{Mahanti}},
  \bibnamefont{and} \bibinfo{author}{\bibfnamefont{C.~G.} \bibnamefont{Olson}},
  \bibinfo{journal}{Phys. Rev. B} \textbf{\bibinfo{volume}{65}},
  \bibinfo{pages}{085108} (\bibinfo{year}{2002}).

\bibitem[{\citenamefont{Aguilera et~al.}(2013)\citenamefont{Aguilera,
  Friedrich, Bihlmayer, and Bl\"ugel}}]{r_13_Aguilera_Blugel_GW_TI_Bi2X3}
\bibinfo{author}{\bibfnamefont{I.}~\bibnamefont{Aguilera}},
  \bibinfo{author}{\bibfnamefont{C.}~\bibnamefont{Friedrich}},
  \bibinfo{author}{\bibfnamefont{G.}~\bibnamefont{Bihlmayer}},
  \bibnamefont{and} \bibinfo{author}{\bibfnamefont{S.}~\bibnamefont{Bl\"ugel}},
  \bibinfo{journal}{Phys. Rev. B} \textbf{\bibinfo{volume}{88}},
  \bibinfo{pages}{045206} (\bibinfo{year}{2013}).

\bibitem[{\citenamefont{Turek et~al.}(2004)\citenamefont{Turek, Kudrnovsk\'y,
  Drchal, and Weinberger}}]{r_04_tkd}
\bibinfo{author}{\bibfnamefont{I.}~\bibnamefont{Turek}},
  \bibinfo{author}{\bibfnamefont{J.}~\bibnamefont{Kudrnovsk\'y}},
  \bibinfo{author}{\bibfnamefont{V.}~\bibnamefont{Drchal}}, \bibnamefont{and}
  \bibinfo{author}{\bibfnamefont{P.}~\bibnamefont{Weinberger}},
  \bibinfo{journal}{J. Phys.: Condens.\ Matter} \textbf{\bibinfo{volume}{16}},
  \bibinfo{pages}{S5607} (\bibinfo{year}{2004}).

\bibitem[{\citenamefont{Carva et~al.}(2007)\citenamefont{Carva, Turek, and
  Kudrnovsk{\'y}}}]{r_07_ct_gamnas}
\bibinfo{author}{\bibfnamefont{K.}~\bibnamefont{Carva}},
  \bibinfo{author}{\bibfnamefont{I.}~\bibnamefont{Turek}}, \bibnamefont{and}
  \bibinfo{author}{\bibfnamefont{J.}~\bibnamefont{Kudrnovsk{\'y}}},
  \bibinfo{journal}{J.\ Magn.\ Magn.\ Mater.} \textbf{\bibinfo{volume}{310}},
  \bibinfo{pages}{2123} (\bibinfo{year}{2007}).

\bibitem[{\citenamefont{Wei et~al.}(2015)\citenamefont{Wei, Lv, Zhang, Yang,
  and Zhao}}]{r_15_Wei_TuningTranProp_Mn_Bi2Se3}
\bibinfo{author}{\bibfnamefont{Z.}~\bibnamefont{Wei}},
  \bibinfo{author}{\bibfnamefont{L.}~\bibnamefont{Lv}},
  \bibinfo{author}{\bibfnamefont{M.}~\bibnamefont{Zhang}},
  \bibinfo{author}{\bibfnamefont{X.}~\bibnamefont{Yang}}, \bibnamefont{and}
  \bibinfo{author}{\bibfnamefont{Y.}~\bibnamefont{Zhao}},
  \bibinfo{journal}{Journal of Superconductivity and Novel Magnetism}
  \textbf{\bibinfo{volume}{28}}, \bibinfo{pages}{2083} (\bibinfo{year}{2015}).

\end{thebibliography}

\begin{thebibliography}{}
\bibitem{vasp} G. Kresse and D. Joubert, Phys. Rev. B {\bf 59}, 1758
(1999).

\bibitem{lmto_cpa} J. Kudrnovsk\'y and V. Drchal,
Phys. Rev. B {\bf 41}, 7515 (1990).

\bibitem{book} I. Turek, V. Drchal, J. Kudrnovsk\'y, M. \v{S}ob, and P.
Weinberger, \textit{Electronic Structure of Disordered Alloys, Surfaces and
Interfaces} (Kluwer, Boston, 1997).

\bibitem{zca} J. P. Perdew and A. Zunger,
Phys. Rev. B {\bf 23}, 5048 (1981); D. M. Ceperley and B. J. Alder,
Phys. Rev. Lett. {\bf 45}, 566 (1980).

\bibitem{pbe} J.P. Perdew, K. Burke, and M. Ernzerhof, 
Phys. Rev. Lett. {\bf 77}, 3865 (1996).

\bibitem{cinane} Y. Li, X. Zou, J. Li, and G. Zhou,
J. Chem. Phys. {\bf 140}, 124704 (2014).

\bibitem{henk} M.G. Vergniory, M.M. Otrokov, D. Thonig, M. Hoffmann, 
I.V. Maznichenko, M. Geilhufe, X. Zubizarreta, S. Ostanin, A. Marmodoro, 
J. Henk, et al., Phys. Rev. B 89, 165202 (2014).

\bibitem{feir} J. Kudrnovsk\'y, F. M\'aca, I. Turek, and J. Redinger,
Phys. Rev. B {\bf 80}, 064405 (2009).

\bibitem{skriver} H. Skriver, {\it The LMTO Method: Muffin-Tin Orbitals 
and Electronic Structure}, (Springer-Verlag, 1984, Berlin). 

\bibitem{zunger} C. Persson and A. Zunger,
Phys. Rev. B {\bf 68}, 035212 (2003).
\end{thebibliography}

%%%%%%%%%%%%%%%%%%%%%%%%%%%%%%%%%%%%%%%%%%%%%%%%%%%%%%%%%%%%%%%%%%%%%%%%%%%%%%%%%%%%%%%%%%%%%%%%%%%%%%%%%%%%%%%%%%%%%%%
%%%%%%%%%%%%%%%%%%%%%%%%%%%%%%%%%%%%%%%%%%%%%%%%%%%%%%%%%%%%%%%%%%%%%%%%%%%%%%%%%%%%%%%%%%%%%%%%%%%%%%%%%%%%%%%%%%%%%%%
%%%%%% SUPPLEMENTAL MATERIAL

\clearpage

\newcommand*{\balancecolsandclearpage}{%
  \close@column@grid
  \clearpage
  \twocolumngrid
}

\newcommand{\beginsupplement}{%
  \setcounter{table}{0}
  \renewcommand{\thetable}{S\arabic{table}}%
  \setcounter{figure}{0}
  \renewcommand{\thefigure}{S\arabic{figure}}%
}
    
\beginsupplement

\part*{Supplemental material}

\section*{\uppercase{Treatment of alloy disorder}}

There are few relevant problems to be addressed when studying
physical properties of present random systems, namely: 
(i) \tcr{a} reliable reproduction of the concentrations trends, 
(ii) \tcr{a} consistent treatment of the transport relaxation times  
or related spectral properties of alloys, 
(iii) \tcr{t}he effect of lattice relaxations and local environment 
effects, in particular in systems with large size mismatch of 
alloy components, and
(iv) \tcr{the} possibility to treat few different impurities in the system.
There exist two methods which address above problems using different
tools\tcr{:}
the supercell (SC) approach, specifically that 
using the \tcr{pseudo}-potential approach (e.g., like the VASP \cite{vasp}), 
and the coherent potential approximation (CPA).  \cite{lmto_cpa,book}
Both methods fulfil well the property (i), as well as the
property (iv), although the CPA is technically more handy in
this case.
The property (ii) is addressed naturally in the framework of the
considered methods only by the CPA, while the SC approach requires 
additional external tools.
On the other hand, the SC method excels in the property (iii) relevant
in the present case of the large atom-size mismatch where lattice 
relaxations/local environment effects can be important.

We will first address the problem of lattice relaxations and their 
relevance for Mn$_{\rm Bi}$ and \MnI{} defects in Bi$_{2}$Te$_{3}$
using the SC-VASP approach.
In the next step we will indicate how the effect of lattice relaxations
can be included approximately in the TB-LMTO-CPA approach which is
then used for transport studies.
As a result, we have found that a combination of both methods, each of 
which represents an alternative approach to the disordered alloy, is 
a useful tool for the study of transport properties of complex alloys.

\subsection{Effect of lattice relaxations: SC-VASP approach}
\label{lr-vasp}

The electronic structure of Mn$_{\rm Bi}$  and \MnI{} is 
studied using the Bi$_{23}$Mn$_{\rm Bi}$Te$_{36}$ (60 atoms) and 
Bi$_{24}$\MnI{}Te$_{36}$ (61 atoms) supercells, respectively. 
This corresponds to the nominal Mn-doping of 4.2\% (1/24).
The van der Waals gap interstitial position was assumed for
\MnI{}.
The projector augmented wave method as implemented in the VASP codes 
\cite{vasp} was used. 
We have tested both the LDA \cite{zca} and the GGA \cite{pbe}
approaches but results are very similar.
Therefore we show below \tcr{the} results for the LDA case \tcr{in order} to compare with
the modified TB-LMTO-CPA method.
The Brillouin zone was sampled by 10$\times$10$\times$2 k-vectors, 
the plane wave cutoff energy was 400~eV, the total energy error was 
better than 0.05 meV per supercell, and the structure was optimized 
\tcr{until} forces acting on each atom were smaller than 2.5~meV/\AA.
During the total energy minimization we have neglected volume
changes due to \tcr{the} doping and fixed volume to that corresponding to
the ideal Bi$_{2}$Te$_{3}$. 
All calculations \tcr{were} scalar-relativistic ones and when 
calculating the densities of states (DOS) the Gaussian 
broadening of 0.1~eV was used. \tcr{The distances between Mn and the first and the second nearest Te atom were relaxed from 3.03 {\AA} to 2.82 {\AA}, and from 3.25 {\AA} to 2.90 {\AA}, respectively.}

\begin{figure}[t]
\includegraphics[width=\columnwidth] {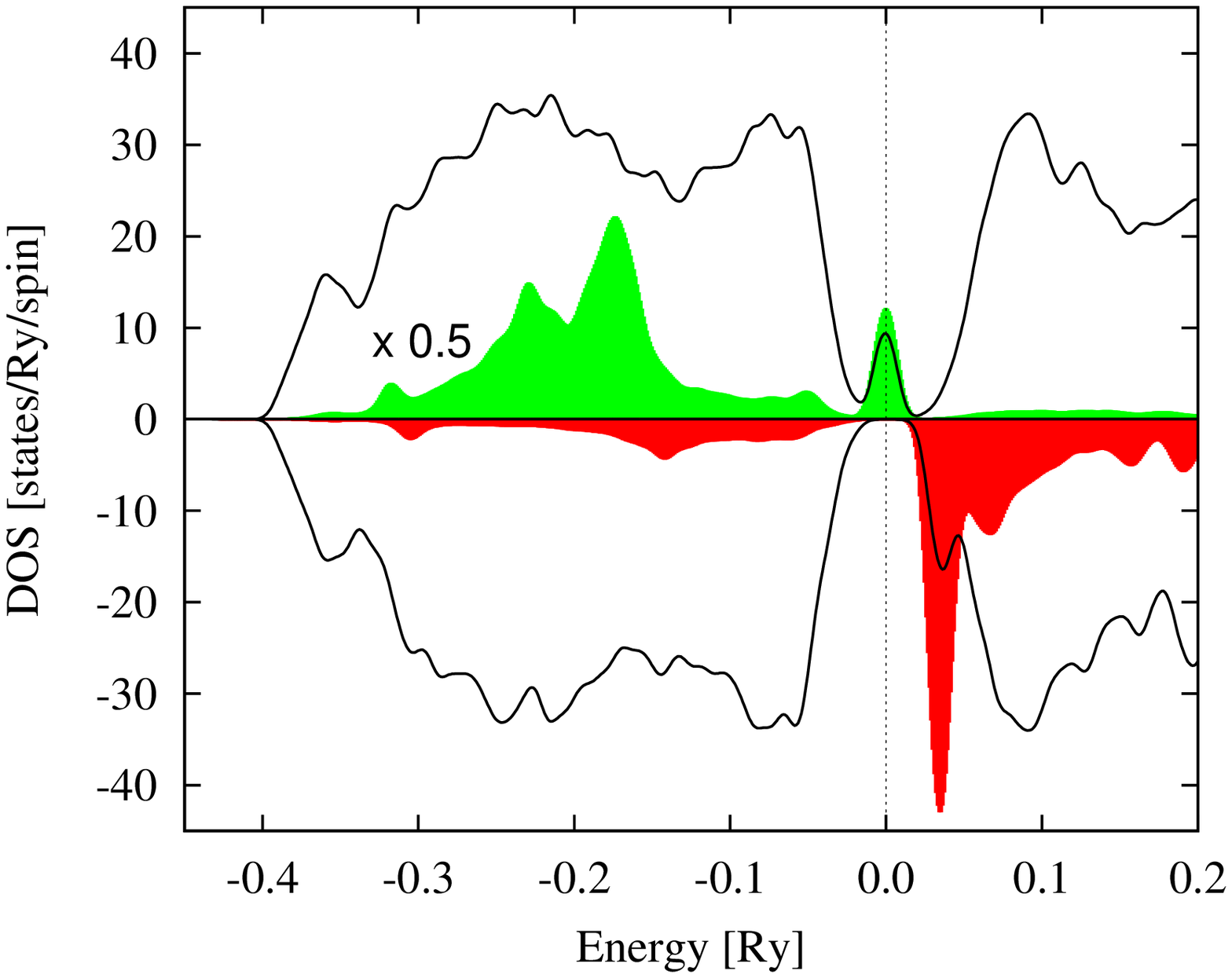}
\caption{ The spin-resolved total (full lines) and Mn-local (coloured
area, factor 0.5) DOS's for the Bi$_{23}$Mn$_{\rm Bi}$Te$_{36}$ 
supercell. The model with lattice relaxations should be compared 
with Fig.~3a of the main text. 
}
\label{sf1}
\end{figure}

\begin{figure}
\includegraphics[width=\columnwidth] {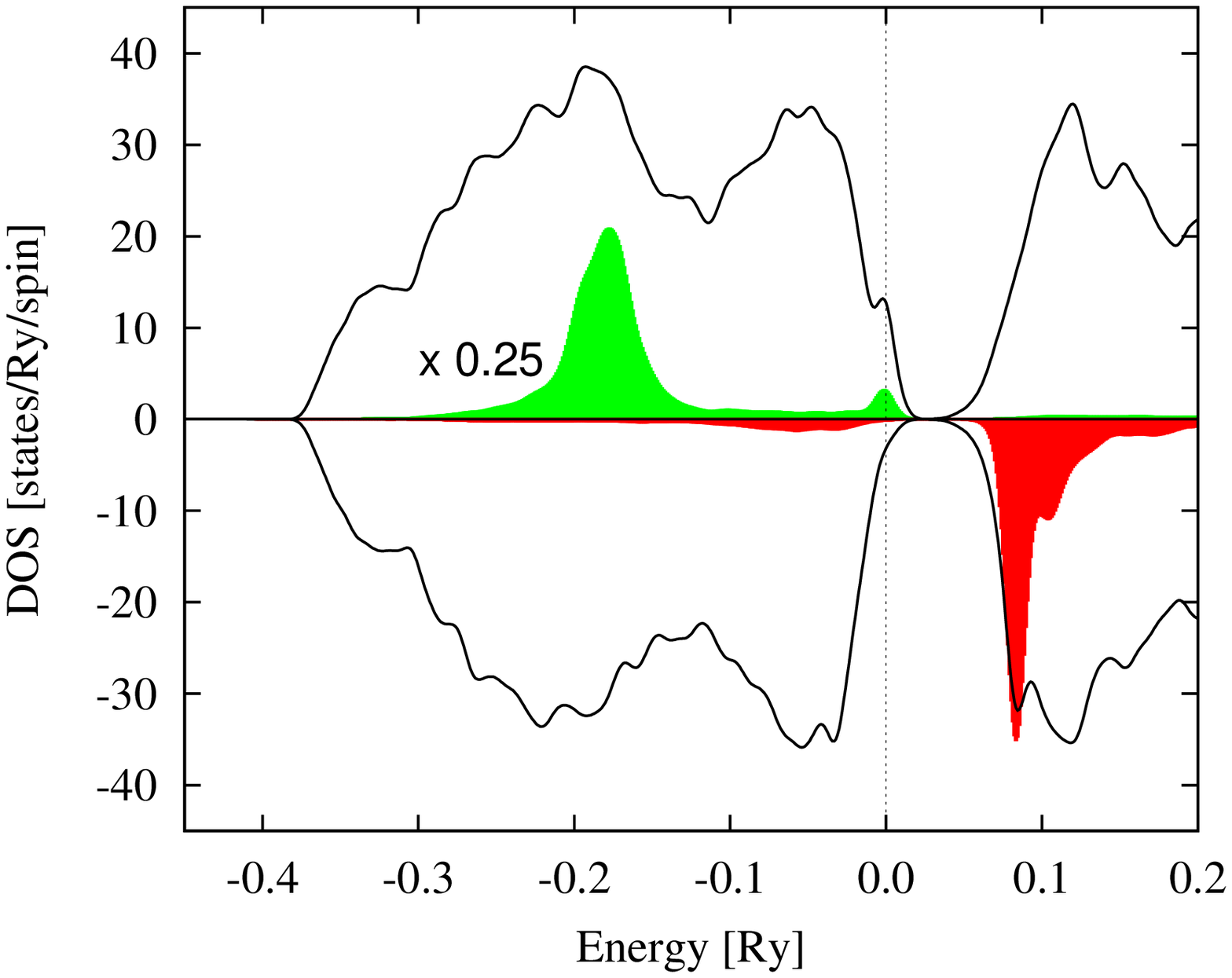}
\caption{ The same as in Fig.~1 but for the unrelaxed 
Bi$_{23}$Mn$_{\rm Bi}$Te$_{36}$ supercell. The factor scaling local
Mn-DOS is 0.25 in this case. }
\label{sf2}
\end{figure}

\begin{figure}
\includegraphics[width=\columnwidth] {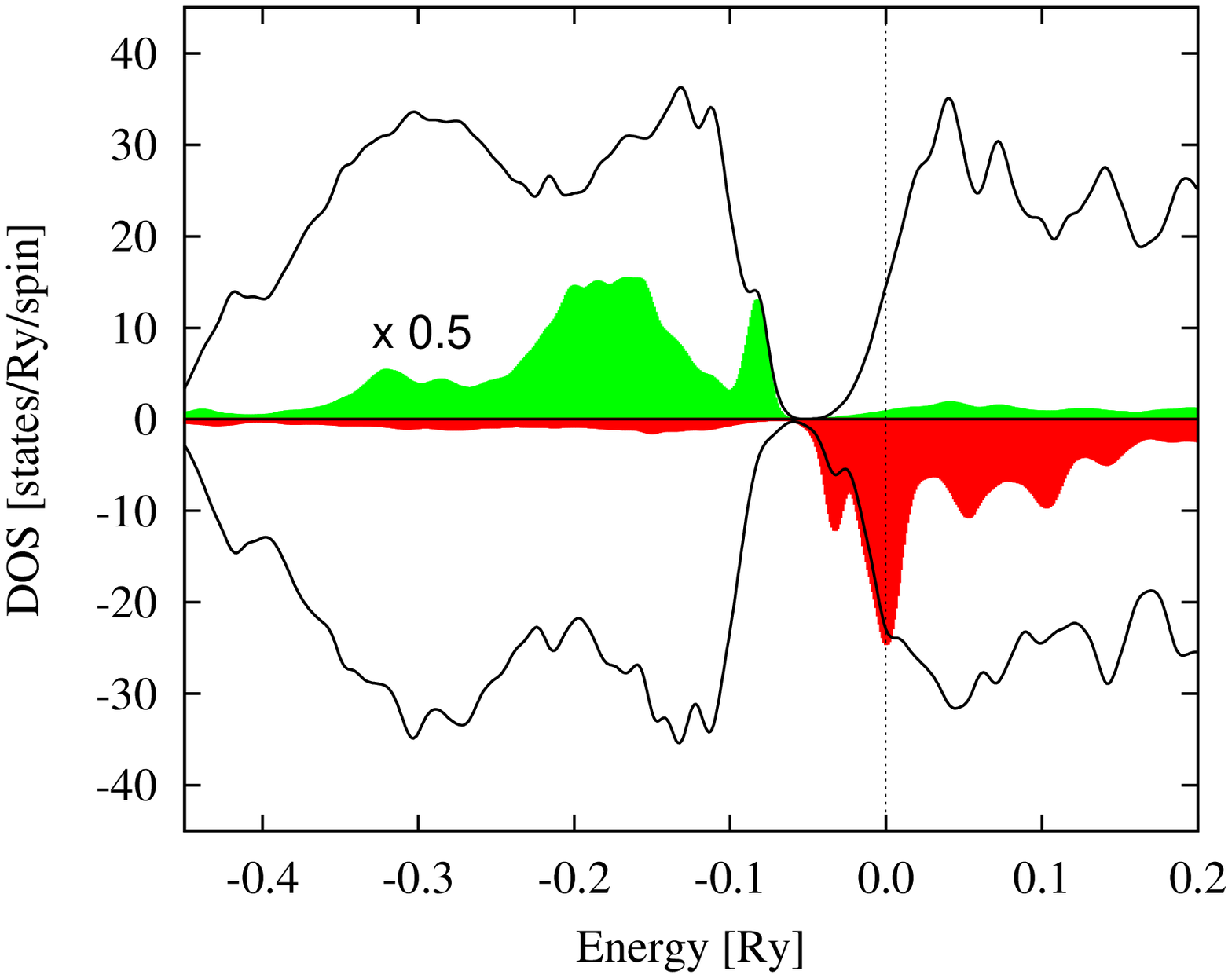}
\caption{ The same as in Fig.~1 but for Bi$_{24}$\MnI{}Te$_{36}$ 
supercell. Negligible effect of lattice relaxations was found in
this case. To be compared with Fig.~4a of the main text.
\label{sf3}
}
\end{figure}

The total and Mn-resolved densities of states (DOS) are shown
for the relaxed (Fig.~\ref{sf1}) and unrelaxed (Fig.~\ref{sf2}) 
Mn$_{\rm Bi}$, respectively, as well as for \MnI{} 
(Fig.~\ref{sf3}).
The most important conclusion is the relevance of lattice 
relaxations for Mn$_{\rm Bi}$ defects.
We observe a clearly pronounced virtual bound state at the Fermi
energy E$_{\rm F}$ for majority states\tcr{,} at the top of the valence 
band.
This state has a pronounced effect on transport properties\tcr{,} which 
depend sensitively on the electronic structure at E$_{\rm F}$.
\tcr{The} corresponding minority state is lying in the unoccupied conduction 
band close to its edge.
Very similar results but assuming the GGA rather than LDA and 
including the spin-orbit coupling were obtained recently in 
Ref.~\onlinecite{cinane}.

The model without lattice relaxations differs significantly due
to \tcr{the} shift of majority/minority Mn-states downward/upward in
energy\tcr{,} which results in a much smaller impurity scattering at
E$_{\rm F}$ and thus \tcr{leads} to \tcr{a} reduced resistivity.
The local Mn-DOS for the unrelaxed case is very similar to that 
obtained in Ref.~\onlinecite{henk} for a closely related case of
Mn\tcr{$_{\rm Sb}$ in Sb$_{2}$Te$_{3}$}.
On the other hand, negligible lattice relaxations were found for
\MnI{} because a small Mn-atom easily accommodates in large 
empty van der Waals site\tcr{,} contrary to the substitutional 
Mn$_{\rm Bi}$ case \tcr{with} the large size mismatch between Mn and Bi.

\subsection{Effect of lattice relaxations: TB-LMTO-CPA}

Methods using the spherical approximation for potentials like,
e.g., the present TB-LMTO-CPA \tcr{ approach,} cannot determine reliably forces 
and thus lattice/layer relaxations.
On the other hand, knowing them from the experiment or full
potential calculations, it is possible to include their
effect on the electronic structure approximately.
 For example, the effect of layer relaxations for fcc-Fe/Ir(001)
overlayer was included approximately in Ref.~\onlinecite{feir}
and results were in a good agreement with full potential 
calculations, in particular the antiferromagnetic ground state 
was obtained correctly only if layer relaxations are included.
It should be emphasized that the method sketched below is not
a substitute for full potential methods but rather a tool how
to capture main effects of lattice relaxations approximately 
and exploit them for the study of complex related physical
properties, like e.g. the transport in the present case. 

The LMTO theory depends on two types of Wigner-Seitz radii,
namely $w^{all}$\tcr{,} which defines the volume of the alloy,
and \tcr{the} local \tcr{radius} $w^{Q}$ corresponding to atomic species $Q$
($Q$=A,B) in an alloy A$_{1-x}$B$_{x}$. 
\cite{skriver,lmto_cpa}
\tcr{The} conventional choice is $w^{Q}$=$w^{all}$ for all atoms.
We have modified this choice locally for $Q$=Mn,Bi in such \tcr{a}
way that the volume of the Bi-sublattice, 
$(1-x) (w^{\rm Bi})^{3} + x (w^{\rm Mn})^{3}$, is preserved.
According to the transformation properties of the LMTO structure
constants \cite{skriver,lmto_cpa} this leads to \tcr{a} corresponding
modification of hopping integrals\tcr{,} and during the selfconsistent
loop \tcr{also} to the change of potential parameters.
It should be noted that in the past we have tested this approach
also for the zincblende GaAs doped with a low concentration of
small nitrogen atoms substituting As-sites, \tcr{a} situation similar 
to that of Mn$_{\rm Bi}$\tcr{, but} for \tcr{a} non-magnetic impurity.
Using this approach, we have correctly reproduced, in agreement 
with the full potential study \cite{zunger}, the downward shift 
of the nitrogen level for the case with lattice relaxations 
included as compared to the unrelaxed case.

We have chosen specifically, $w^{\rm Mn}$=2.8~a.u., typical
native value for solid state Mn, and $w^{\rm Bi}$ accordingly.
It should be noted, that $w^{Q}$ radii are used for the
solution of the LDA Schr\"odinger equation. \cite{book}
Corresponding DOS\tcr{'s} (see Fig.~3a and Fig.~4a in the main text)
demonstrate that the main features obtained in the SC-VASP
calculations (see Fig.~\ref{sf1} and Fig.~\ref{sf3} in this Supplement)
were reasonably well reproduced.
Small differences can be ascribed to the fact that the finite Gaussian 
broadening inside the Brillouin zone was used for SC-VASP\tcr{,} as contrasted 
\tcr{to the} anisotropic damping due to the CPA.
No modifications were done for the case of Bi$_{\rm Te}$ and
Te$_{\rm Bi}$ antisites.

% set reference counter
\makeatletter
\apptocmd{\thebibliography}{\global\c@NAT@ctr 47\relax}{}{}
\makeatother

\end{document}